\documentclass[manuscript]{acmart}
\AtBeginDocument{%
  \providecommand\BibTeX{{%
    \normalfont B\kern-0.5em{\scshape i\kern-0.25em b}\kern-0.8em\TeX}}}

\setcopyright{none}

\usepackage{booktabs}
\usepackage{subfigure}
\usepackage{changepage}

\newcommand{\isac}{\textsc{GP-TSM}}
\newcommand{\ngp}{\textsc{NGP-TSM}}
\newcommand{\static}{\textsc{HITL-GP-TSM}}
\newcommand{\wf}{\textsc{WF-TSM}}
\newcommand{\interactive}{\textsc{HITL-GP-TSM-Interactive}}
\newcommand{\control}{\textsc{Control}}

\definecolor{GRAY1}{gray}{0.2}
\definecolor{GRAY2}{gray}{0.4}
\definecolor{GRAY3}{gray}{0.6}

\begin{document}
\title{An AI-Resilient Text Rendering Technique for Reading and Skimming Documents}%
\author{Ziwei Gu}
\email{ziweigu@g.harvard.edu}
\affiliation{%
  \institution{Harvard University}
  \country{USA}
}

\author{Ian Arawjo}
\email{ian.arawjo@gmail.com}
\affiliation{%
  \institution{Harvard University}
  \country{USA}
}

\author{Kenneth Li}
\email{ke_li@g.harvard.edu}
\affiliation{%
  \institution{Harvard University}
  \country{USA}
}

\author{Jonathan K. Kummerfeld}
\email{jonathan.kummerfeld@sydney.edu.au}
\affiliation{%
  \institution{University of Sydney}
  \country{Australia}
}

\author{Elena L. Glassman}
\email{glassman@seas.harvard.edu}
\affiliation{%
  \institution{Harvard University}
  \country{USA}
}

\begin{abstract}
Readers find text difficult to consume for many reasons. Summarization can address some of these difficulties, but introduce others, such as omitting, misrepresenting, or hallucinating information, which can be hard for a reader to notice. One approach to addressing this problem is to instead modify how the original text is rendered to make important information more salient. We introduce Grammar-Preserving Text Saliency Modulation (\isac{}), a text rendering method with a novel means of identifying what to de-emphasize. Specifically, \isac{} uses a recursive sentence compression method to identify successive levels of detail beyond the core meaning of a passage, which are de-emphasized by rendering words in successively lighter but still legible gray text. In a lab study (n=18), participants preferred \isac{} over pre-existing word-level text rendering methods and were able to answer GRE reading comprehension questions more efficiently.
\end{abstract}



\begin{teaserfigure}
\centering
\includegraphics[width=0.9\textwidth]{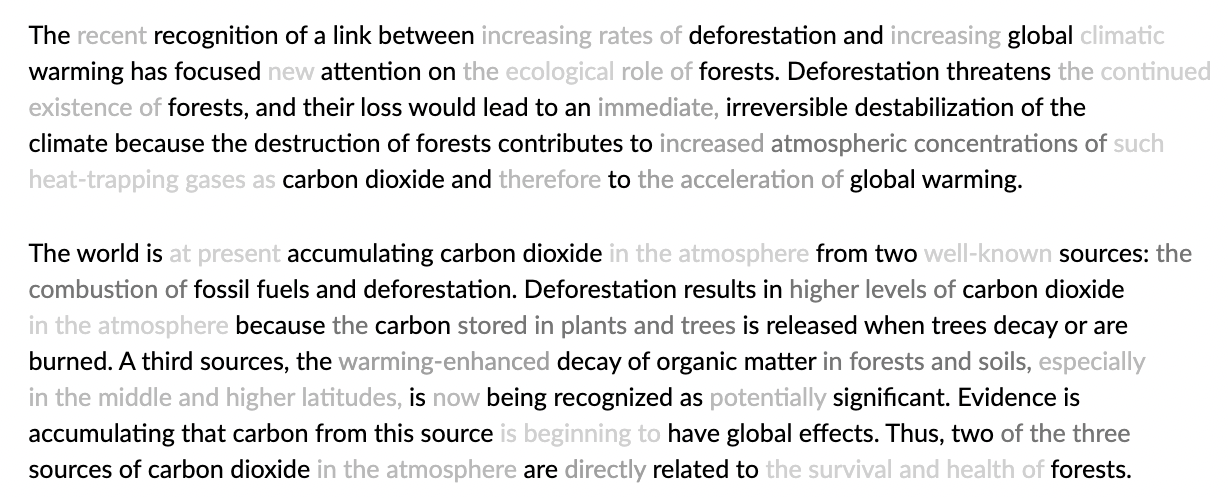}
\caption{The \isac{} rendering technique applied to two paragraphs from a passage from the GRE reading comprehension test.
The lighter the text color that each word is rendered in, the earlier it was cut in the backend's recursive sentence compression process. The darkest subset of text can be read as grammatical sentences that preserve as much of the semantic value of the original document as possible, and every successive level of lighter text can be added to these darkest sentences---adding detail without modifying their grammaticality.}
\label{fig-interface}
\end{teaserfigure}

\maketitle

\section{Introduction}

Readers can find text difficult to consume for a variety of reasons related to the author(s)' choices and the readers' skills and context. First and foremost, there may be a large volume of text relative to the time and attention the reader is willing or able to set aside to read it. In addition, sentences may be long, have a complex structure, and/or have ambiguous parses, e.g., `garden-path sentences'~\cite{zipoli2017unraveling}. The subset of the language used by the author(s) may not have a high degree of overlap with the reader's sight vocabulary~\cite{dolch1936basic}. Finally, the reader may be still learning how to read in that language, or the reader may have cognitive differences or conditions that impede reading. 

Automated text summarization techniques, including but not limited to crowd-powered systems~\cite{bernstein2010soylent}, prompting large language models (LLMs)~\cite{zhang2023benchmarking}, and other AI technologies, can address a subset of these difficulties, i.e., the resulting text may be shorter, with simpler sentence structures and fewer unusual words~\cite{mani2001automatic}. However, unless there is information within the original document that is truly redundant, the result is a lossy representation of the original document, regardless of whether the process is abstractive\footnote{A method of summary generation where the system creates a condensed version of the source text using novel sentences. It rephrases the original content to produce a coherent summary, potentially introducing new words and structures.} or extractive\footnote{A method of summary generation where the system selects and extracts whole sentences or fragments directly from the source text to construct the summary. It does not modify the original content but rather curates important segments to form the summary.}.

Specifically, automated summarization methods can introduce multiple types of errors, i.e., ``crimes'' of omission, hallucination, and misrepresentation. Specifically, they may judge some details as insufficiently relevant and omit them when they are actually crucial to the reader, given their particular knowledge, context, preferences, values, and task. They may introduce false or irrelevant information that is not derived from the original text, a phenomenon known as confabulation\footnote{In psychology, the term ``confabulation'' describes ``honest lying''\cite{gilboa2002cognitive}. The term ``hallucination'' may be more popular among computer scientists when referring to generative AI.} or hallucination~\cite{maynez2020faithfulness, ji2023survey, zhao2020reducing}. And when summarizers paraphrase or choose what subset of the original text to preserve in the summary, they risk shifting the resulting meaning further away from the original text than the reader would accept, given their goals and context, \textit{if they knew} (i.e., misrepresentation). Every reader may have their own context, tasks, and tolerances, and no single summary can be perfect for all. While personalised summarization has been studied for over a decade~\cite{yan-etal-2011-summarize}, it still relies on a coarse characterizations of users~\cite{xu-etal-2023-pre-trained} and there is still much about a reader's interests, context, and task in the moment that will be unobservable.

All possible instances of these errors in AI-generated summaries are impossible for readers to notice unless they also read the entire original document. Errors of hallucination tend to look plausible at a glance, errors of omission leave nothing to be noticed in the summary itself,\footnote{This interface challenge is analogous to how users cannot recognize the false positives of spam detection algorithms just by looking at their inbox, because the decisions are made silently, leaving no trace in the inbox itself; users have to explicitly look at every email their spam folder to exhaustively find such false positives.} and 
errors of misrepresentation are undetectable unless readers also read all the relevant portions of the original document (which may be the entire document).
Recovering from these AI errors is hard because readers have to first notice AI choices that may or may not reflect one of these errors, and have enough context to judge whether or not they are one of these errors, which are pre-requisites to the previously proposed human-AI interaction design guidelines~\cite{amershi2019guidelines} ``support efficient dismissal'' and ``support efficient correction.'' We call an interface that supports users in noticing, judging, and recovering from AI errors like these \textit{AI-resilient}. 

One potentially AI-resilient alternative approach to automated summarization is to instead modify the visual attributes of the original text to support faster reading of the original document (skimming). We propose and evaluate such an approach, which we call Grammar-Preserving Text Saliency Modulation (\isac{}). Its novelty comes from the computational method used to determine which words in the original text to de-emphasize---and by how much. Specifically, \isac{} uses a recursive sentence compression method to identify successive levels of detail beyond the core meaning of a passage, which are de-emphasized by rendering words with successively less opacity, e.g., lighter and lighter but still legible gray text when black text is on a white background. The lighter the text color that each word is rendered in, the earlier it was cut in the backend's recursive sentence compression process. We describe the approach as "grammar-preserving" because the subset of each sentence at any minimum level of opacity the reader chooses to read, remains grammatical---which supports a more natural flow of reading. A formative study in which \isac{} was semi-automated (with a human in the loop) validated the value it would provide if fully automated.

Prior text rendering methods have computed a variety of functions over words and sentences within a document (from unigram frequency~\cite{brath2014using, 10.1145/1753846.1754093} to neural-network-based semantic similarity~\cite{yang2017hitext, fok2023scim})  and reified the results of that computation into a variety of visual attribute modifications including font attributes~\cite{wallace2022towards, brath2014using, 10.1145/1753846.1754093, stoffel2012document, parra2019analyzing} and background color~\cite{fok2023scim, wecker2014semantize, yang2017hitext}. In particular, two previously published ideas that were presented \textit{without evaluation} propose helping readers skim by reifying unigram frequency using font weight as described by~\citet{brath2014using} or opacity, i.e., \textsc{QuickSkim}~\cite{10.1145/1753846.1754093}. In our final user study, we compare font opacity modulated by \isac{} to font opacity modulated by unigram frequency (as a control condition we call \wf{}). 

A within-subjects user study (N=18) demonstrates that the final design of \isac{} not only helps readers complete non-trivial (GRE) reading comprehension tasks more efficiently, it is also strongly preferred over font opacity modulated by unigram frequency (\wf{}). 

In summary, we contribute:
\begin{itemize}
  \item The design and implementation of \isac{}, a recursive sentence-compression-based text rendering method that supports reading and skimming
  \item A formative within-subjects user study that demonstrates the value of GP-TSM's text rendering strategy---using a semi-automated sentence-compression backend
  \item A summative within-subjects user study that (1) demonstrates the benefits of the fully automated \isac{} relative to prior text rendering methods and (2) collects evidence that GP-TSM's preservation of grammar at every level of successively grayed text is key.
\end{itemize}

\section{Related Work}
Made possible by the capabilities of large language models (LLMs), \isac{} is an extension of a range of prior work on text summarization and text rendering intended for reading, skimming, and information retrieval support. In this section, we seek to contextualize \isac{} within the broader narrative of natural language processing (NLP) and the foundational role of earlier research on text rendering and reading and skimming support systems.

\subsection{Reading and Skimming}

Natural language documents pervade people's everyday lives, and reading them can require non-trivial mental effort~\cite{rayner2012psychology, janicke2014designing, janicke2015close}. In particular, long, complicated sentences can be hard to understand~\cite{zipoli2017unraveling}. Studies from the American Press Institute~\cite{dubay2004principles} show that when a document's average sentence length is 14 words, readers understand more than 90\% of what they are reading. At 43 words, comprehension drops to less than 10\%. But long complex sentences remain a common occurrence. For example, recent education research articles written in English had an average sentence length of 24.7 words~\cite{deveci2019sentence}, similar to news text and biomedical text, which average 24.8 and 24.5 words per sentence respectively in standard corpora \cite{kummerfeld-etal-2010-faster}.

Given the cognitive effort reading requires, readers frequently resort to skimming, which is a rapid, selective, and non-linear form of reading~\cite{agosto2002bounded}. Eye tracking studies~\cite{duggan2011skim, pernice2014people} validate that such behavior is extremely common. However, multiple studies have suggested a significant trade-off between reading speed and comprehension ~\cite{masson1982cognitive, masson1983conceptual, tashman2011active, rayner2016so}. In addition, skimming, a skill that takes time to learn and employ effectively, requires strategy and attention~\cite{duggan2009text}. In particular, when skimming in unfamiliar contexts, readers tend to struggle to stay focused, miss key information, and lack confidence in their understanding~\cite{duggan2006much, yi2014qndreview, mangen2016evolution}. Studies have shown that the comprehension of important and unimportant information from a text was equally degraded by an increase in reading rate~\cite{carver1984rauding, dyson2000effects, kiwan2000effects}.

\subsection{Text Summarization Methods}
Text summarization can be either extractive or abstractive. Extractive summarization selects a set of text segments from the original document(s) and combines the segments to form a summary. Note that in these approaches, the summary is entirely composed of verbatim content, i.e., words have been removed but none have been added.
The earliest extractive systems selected a set of sentences~\cite{luhn-58}.
More recent work has also compressed and/or merged the sentences that are selected \cite{lin-2003-improving}. A range of modeling~\cite{dong-etal-2018-banditsum,luo-etal-2019-reading,jia-etal-2020-neural,xu-etal-2015-extractive,jadhav-rajan-2018-extractive} and learning~\cite{zhong-etal-2020-extractive,wang-etal-2019-self-supervised,zhang-etal-2018-neural,narayan-etal-2018-ranking} methods have been explored.

One drawback of the extractive approach is that it can be difficult or impossible to \emph{concisely} capture meaning while only using verbatim content; this is in contrast to abstractive summarization, which generates novel sentences to capture the essence of the content~\cite{allahyari2017text, zhang2017sentence, zhong2022unsupervised}. Abstractive summarization can be more flexible, concise, and human-readable. Historically, extractive summarization was more successful in terms of accuracy and coherence, but recent improvements in natural language generation using LLMs has made abstractive summarization effective and popular~\cite{widyassari2022review}. 

For our specific application, only extractive summarization is suitable. Our goal is to modulate the saliency of words in the \emph{original text} so that users can easily bypass certain words during skimming while maintaining an uninterrupted reading flow. This goal aligns with a specific family of extractive summarization known as sentence compression, or compressive summarization. While traditional extractive summarization predominantly involves selecting whole sentences, compressive summarization aims to select the shortest subsequence of words \emph{within} an sentence that yields an informative and grammatical sentence~\cite{martins2009summarization}. This framework allows for a more concise representation of the original content while retaining the essence of its meaning. Various techniques have been developed within this framework~\cite{berg2011jointly, filippova2015sentence, desai2020compressive, xu2019neural, klerke2016improving, mendes2019jointly, zhao2018language}. \textit{Notably, our approach introduces a novel feature---a recursive process that generates multiple nested levels of compression, in which information is captured at varying degrees of detail.} In contrast, while there has recently been some work on generating a set of summaries that vary in detail, it has been abstractive, with content at each level that does not overlap~\cite{zhong-etal-2022-unsupervised}. 

A range of technologies have been applied to summarization, from traditional NLP techniques~\cite{mihalcea2004textrank, filippova2008dependency, allahyari2017text, berg-kirkpatrick-etal-2011-jointly} to large language models (LLMs)~\cite{zhang2023extractive, zhang2023benchmarking, zhang2023summit} and even crowd-powered methodologies~\cite{bernstein2010soylent}.
The recent improvements in LLMs has significantly increased the quality of summarization methods, including compressive summarization. The summaries they produce are better in terms of coherence, grammaticality, and coverage of critical content~\cite{fang2023chatgpt, wu2023chatgpt}.

There are a variety of systems that employ summarization within their enhanced reading environments. Specifically, many systems add abstractive and/or extractive summaries to give the reader additional, shorter, possibly simpler text that augments the original content. For example, 
Paper Plain~\cite{august2022paper} uses AI models to generate abstractive summaries of each section of a medical paper which is intended to make the science literature more approachable to healthcare consumers. Marvista~\cite{chen2022marvista}, a human-AI collaborative reading tool, employs an extractive strategy in the "before reading" phase and automatically chooses a summative subset of text for users based on their time budget and questions they want to answer. Marvista then uses AI-generated abstractive summaries to help readers review and recall important information from articles in the "after reading" phase. \textsc{newsLens}~\cite{laban2020framework} describes a quote-extraction based summary using entity extraction and dependency trees to complement news headlines and represent potentially important details from the rest of the article.

However, regardless of the method used, both abstractive and extractive summarization can introduce significant changes in semantic meaning, e.g., through misinterpretation of the input or unintended meanings of the output~\cite{allahyari2017text, zhao2020reducing}. To mitigate this lossy nature of summarization methods, our approach supports reading and skimming by adjusting the way text is rendered, keeping all text available to the user to enable recovery from AI errors.

\subsection{Text Rendering Modulation Methods}

Extensive research has been conducted on text rendering methods to enhance readability, with a significant focus on font attributes. Prior studies have demonstrated that reading performance can improve when using a font that is individually optimal for a reader~\cite{bernard2001effects, dobres2016utilising, chatrangsan2019effect}. However, there isn't a universally ideal font suitable for all readers, and a reader's optimal font may not always align with their preferred choice~\cite{wallace2020accelerating, wallace2022towards}. Building upon this line of research, a machine-learning-based model named FontMART has been developed to predict the font that enable the fastest reading speed for an individual~\cite{cai2022personalized}. Our work aims to complement this research in readability by focusing on making key information more \textit{salient}, thereby facilitating both focused reading and efficient skimming.

Modulating text saliency is a widely studied aspect of textual information representation. This technique modifies the visual attributes of text to promote words of interest and guide readers' attention, making pertinent information more perceptible and thereby enhancing comprehension and the user experience~\cite{itti2007visual, brath2020visualizing}. We adopt the term ``saliency'' based on its definition (a ``bottom-up, stimulus-driven perceptual quality which makes some items stand out from their neighbors'')~\cite{itti2007visual}, and its use in augmented reality~\cite{veas2011directing, sutton2022look}, computer vision~\cite{li2014visual, cedillo2018spatiotemporal}, and cognitive science~\cite{li2002saliency, hadizadeh2016saliency}. 
  
A range of visual strategies have been introduced to promote text saliency in digital reading environments. Brath's \textit{Visual Encoding Pipeline Extended for Text}~\cite{brath2020visualizing} describes how different types of textual data, including the literal text itself, can be mapped to visual attributes, and then drawn as marks in a layout. \citet{brath2014using} describes the varied use of typography, or font attributes such as bold, italic, font size and case, on an individual word level to convey information in a way that is intended to facilitate skimming and text analysis; unlike \isac{}, multiple visual attributes of text were simultaneously used rather than just text color and no controlled studies are presented that evaluate their effectiveness. \citet{10.1145/1753846.1754093} propose a prototype called \textsc{QuickSkim} that assigns a lower text opacity value to non-content words like articles and junctions if skimming is detected from eye tracking. Stoffel et al.~\cite{stoffel2012document} focus on the size of individual words and create thumbnails by retaining a readable font size of interesting text while shrinking less interesting text. Strobelt et al.~\cite{strobelt2015guidelines} surveyed and tested the effectiveness of nine common text highlighting techniques, including various font attributes such as font color, font style and font weight, both individually and in combination. However, their studies involved tasks such as visual interference examination and visual conjunctive search, while our studies focus on reading and skimming. Similarly, Parra et al.~\cite{parra2019analyzing} explores multiple types of encoding of information on documents, including font size, font luminance, and background color lumination/ saturation, but for a different purpose: visualizing neural attention directly on text.

Color as a tool for emphasis and differentiation has been employed in multiple systems~\cite{yang2017hitext, fok2023scim}. For example, \textsc{Scim}~\cite{fok2023scim} introduced a color-coded system to label different types of (entire) sentences in scientific articles. \textsc{HiText}~\cite{yang2017hitext} uses sentence highlighting at various saliencies, where the saliency of the highlight corresponds to the position of the sentence in a list ranked from most to least predicted importance. Semantize~\cite{wecker2014semantize} conveys the sentiment of specific sentences or paragraphs of a document by rendering them with different visual attributes, i.e., background color to represent polarity of emotion; italics for sentences predicted to be subjective; underlines for ``emotion'' words (green or red depending on predicted emotion); font weight to render predicted intense or diminutive words; and font size and spacing modulated by predicted reading grade level. 

\isac{} is solidly within this existing text rendering modulation tradition. \isac{}'s primary contribution is in its method for computing \textit{what} to de-emphasize. Our approach, recursive sentence compression, enables the visual distinction of \textit{multiple nested grammatical subsets of sentences}. As a result, our novel method of computation also presents a new \textit{scope} of visual attribute modulation.

There are multiple existing computational methods for determining text saliency. Both \citet{brath2014using} and \citet{10.1145/1753846.1754093} weight words based on English language word frequency or document-level unigram frequency. Such word-frequency-based methods are optimized for highlighting unique words, but do not take into account the core meanings of a text document or the relationship between words within the same sentence, like \isac{} does. More recently, \textsc{HiText}~\cite{yang2017hitext} uses a neural-network based AI model to rank sentences according to predicted importance. Like \textsc{HiText}, \isac{} uses an AI model, but for a different purpose, i.e., performing sentence compression recursively.

Many systems use text rendering modulation methods, highlights, and/or annotations to visualize their analyses of text in-place. For example, \textsc{Scim}~\cite{fok2023scim} helps readers skim scientific articles by highlighting sentences about different key aspects of a paper using different colors, with a density configurable by readers. \textsc{VarifocalReader}~\cite{koch2014varifocalreader} automatically annotates and highlights text segments in the detail layer and uses the opacity of the annotation highlight to indicate the confidence value from its support-vector-machine-based active learning component. TextViewer~\cite{correll2011exploring}, built for literary scholars, renders text with colored underlines to denote text that has been tagged, with the saturation value of the underline corresponding to the absolute value of the tag weight. \isac{} is versatile and easily integrable, making it suitable for use in these and other contexts, much like the other text rendering methods discussed above.

\section{\isac{}}
Building upon the insights gleaned from our review of the challenges of reading and skimming, existing text summarization methods, and existing text rendering modulation methods, this section describes the design and implementation of our proposed text rendering method---\isac{}. We provide a comprehensive overview of our design goals, the design space, and our design process, followed by explanations of \isac{} and its implementation.

\subsection{Design Goals}
We aspired to design a text rendering interface that alleviates some of the cognitive demands of reading, skimming, or performing information retrieval on natural language documents---particularly those with long, complicated sentences---without compromising the integrity of the original content. From our design explorations, we decided that an effective interface toward that objective should have the following requirements:

\begin{enumerate}
\item \textbf{Remain faithful to the original text}. The system should not automatically reword or add new words or phrases to the original text. It should preserve the original text, while rendering it in a way that aids reading, skimming, or information retrieval. This principle of preserving the integrity of the original content is also a primary design goal of a previously developed tool, Doccurate~\cite{sultanum2018doccurate}, which was developed for a specific domain where precise wording is critical, i.e., healthcare. 
\item \textbf{Integrate seamlessly into existing reading experiences}. The system should complement and not interfere with the existing digital reading workflow that people are already used to. It should provide all the functionalities in the same view, minimizing the overhead of mode and context switching. This principle also guided HiText~\cite{yang2017hitext}; they called this goal ``ergonomic unobtrusiveness''. 
\item \textbf{Support reading at multiple levels of detail}. The system should help users navigate the full complexity of a text, shifting focus seamlessly between different levels of semantic coverage, or granularity~\cite{mulkar2011granularity, zhong2022unsupervised}, from the big picture to the fine details. It should allow users to decide how much detail they want to read and, in case they want a closer read, enable them to do so without requiring any extra action on the users part, e.g., to reveal them. Lower levels of detail are not hidden not only to save the user from having to take an action to reveal them, but also so that users do not have to guess whether they care enough about any hidden detail in order to decide whether to take that action to reveal them.
\item \textbf{Support skimming without interrupting flow.} 
The system should improve skimming of text while minimizing the impact on the user's natural reading flow. In particular, as much as possible, it should avoid presenting users with salient text that is unparsable as a coherent thought, i.e., a complete sentence rather than a phrase or sentence fragment. 
\item \textbf{Be resilient to AI errors by enabling the reader to (a) notice (b) have enough context to judge and (c) easily recover from automated decisions they disagree with.} If the system makes an (automated) judgement call that is inappropriate given the reader's values, preferences, knowledge, context, or task, the reader should be able to recognize that without taking any additional action beyond looking at the interface itself, and proceed without being negatively affected by it. This design goal adds the critical observation that \textit{noticing} an AI's choice and \textit{having enough context to accurately judge} an AI's choice as deserving of preservation or dismissal \textit{are pre-requisites to the previously proposed human-AI interaction design guidelines~\cite{amershi2019guidelines} ``support efficient dismissal'' and ``support efficient correction.''} For example, hidden details critical to a reader but judged insufficiently important to keep by an AI cannot be noticed by virtue of them being hidden, unless the user is lucky enough to take an action to reveal them and discover that they disagreed with the AI's judgement.
\end{enumerate}

\subsection{Design Process}
\begin{table*}[t]
\small
\centering
\begin{tabular}{l|l|l|l|l}
               &                                 &                                     & Summary & Extractive Summarization \\
Text Attribute & Scope of Attribute Modification & Interaction technique               & Scope   & Model                    \\ \midrule
highlight opacity & character         & clicking (a button)             & sentence                     & TF-IDF                               \\
highlight color & word       & clicking (a carousel)                 & \textbf{paragraph}                    & constituency tree analysis                    \\
\textbf{font opacity} & phrase        & dragging a slider handle  & document                     & dependency tree analysis                      \\
font hue  & 
\textit{(novel)} \textbf{nested grammatical}         
& gesturing (pinch to zoom)         & corpus    & linear programming                   \\
font size & 
\textbf{subset(s) of the sentence}            
& pressing keys on a keyboard          &                              &  Latent Semantic Analysis \\
font weight  &  entire sentence     & scrolling with a mouse         &                              &  autoencoder  \\
font width  &        & swiping on a touchscreen         &                              & \textbf{Large Language Model}   \\
oblique &         &  \textbf{toggling rendering on/off}          &                              &    \\
typeface &         &         &                          &    \\
underline  &        &         &                          &    \\
case     &                &                              &      \\  
background color     &                &                              &      \\  
\end{tabular}
\caption{The design space we explored for interfaces that support reading, skimming, and/or information retrieval, including 5 main parameters and alternative values for each parameter. It is influenced by \citet{brath2020visualizing}'s purpose-agnostic \textit{Visualization Encoding Pipeline Extended for Text}. The bolded items describe the final design of \isac{}.}
\label{tab:design_space}
\end{table*}

The design of \isac{} is the result of numerous iterations. To thoroughly explore the design space of skimming support interfaces, we started off by delineating a diverse set of key dimensions and the potential options for each (as detailed in Table~\ref{tab:design_space}).
To ground our explorations, we constructed prototypes within a browser application, each encompassing different combinations of candidate text attributes, text attribute modulation scopes, interaction techniques, computation scopes, and methods of computing what will be rendered with those text attributes. This approach helped us explore key points within the design space without necessarily implementing all possible feature combinations.
In the rest of this section, we describe the process that led to \isac{}'s final features and pivotal design choices, using language consistent with Brath's textbook on textual visualizations~\cite{brath2020visualizing}.

\subsubsection{Text Attribute to Modify}

Similar to \textsc{QuickSkim}~\cite{10.1145/1753846.1754093}, we chose to modulate opacity---which, for black text on a white background, is equivalent to choosing font colors on a gray scale---as opposed to other alternatives like background color or stylistic indicators such as italics, typefaces, and underlines, because of our interest in minimising visual distraction. Modulating opacity allows for a graded emphasis on text without disrupting the visual cohesion of the paragraph, offering a smooth reading experience. Since it is a continuous feature, it can be modulated to varying degrees to differentiate multiple levels of detail as well.

We also found opacity modulation to be a generally intuitive mapping of meaning for users. For example, with black text on a white background, lighter text that has less contrast with the background denotes detail, while darker text signifies criticality. 
Regardless of the text and background colors, modulating opacity allows the text containing the details to have less and less contrast with the background behind it---``fading away'' or moving back ``into the background''. We tried altering other font attributes, such as font hue or width, but found that their meaning was less clear to participants in early pilot studies. 

To fulfill our design goals, we ensure that even the least opaque text is still legible, i.e, consistent with guidelines on contrast ratios provided by WCAG (Web Content Accessibility Guidelines)~\cite{caldwell2008web}, so that words that the computational method deems to be details are not hidden. (If they were hidden, they would be unnoticeable by a reader who needed them given their context.) Nevertheless, our design may still pose challenges for certain groups of people, which we further discuss in the Discussion section.

Deciding to use opacity instead of similar attributes, like font weight or bolding, was also difficult. Bold text inherently demands attention, drawing the reader's eye immediately to those words or phrases~\cite{brath2020visualizing}. While this is effective for emphasizing certain sections, it is contradictory to our goal of de-emphasizing or suggesting skippability. On the other hand, fading out text provides a more subtle indication of detail level without aggressively diverting the reader's focus. Moreover, bold can be visually overpowering and might create visual fatigue over extended reading periods, especially in documents with frequent emphasis changes~\cite{keyes1993typography}. By contrast, the less obtrusive fading method enables a more balanced and smooth reading flow---if the computational method chosen makes relatively ``smooth'' predictions across sequential words in a sentence.

\subsubsection{Scope of Attribute Modification} Our chosen scope of attribution modification could be defined as \textit{word}, but words or even phrases, unlike prior work, are not considered independently outside of their sentence. Instead, the attribute of \textit{grammatical subset(s) of each sentence} are modified, which reveal as many levels of successively smaller detail as the method of computation identifies during its recursive sentence compression process. This is aligned with the design goal of supporting skimming without interrupting flow; one can skim at any minimum level of opacity (skipping over, if one chooses, the words between, which are faded even more) and still be reading coherent sentences that preserve as much of the semantic meaning of the original as possible.

\subsubsection{Interaction techniques, if any}

Our choice to focus on desktop computers instead of mobile devices was influenced by studies that found improved memory and better performance when people read on desktop computers compared with mobile devices~\cite{alrizq2021analysis, shrestha2007mobile}. We conjecture that desktop environments may offer a more conducive setting for extensive reading, particularly for longer and more complex documents, since a larger screen displays more information in a single view without requiring frequent scrolling or zooming. Therefore, we concentrated solely on desktop interactions involving the mouse and keyboard, setting aside interaction alternatives such as swiping and gesturing. 

After implementing and piloting a variety of desktop interactive techniques, including sliders, carousels, and mouse scrolling mechanisms for transitioning between hiding/revealing different levels of information granularity, our final design of \isac{} can just be turned on and off, by keyboard or mouse. Our rationale for this is two-fold: rooted in both our design goal of ensuring seamless integration into existing reading workflows and allowing readers to notice and recover from automated decisions they disagree with. Hiding information interferes with the latter goal, and preliminary studies indicated that the choice interfered with the former goal as well. These preliminary studies, which included a mouse-scrolling feature, suggested that such interactive elements could inadvertently disrupt reading, diverting user attention from the primary task of comprehension. We also observed that other interaction methods could overcomplicate the system, which intimidated some users due to the steeper learning curve. By designing a system that automatically determines and displays text saliency without demanding active user adjustments, we aim to reduce the cognitive load required for reading, an already demanding task.

\subsubsection{Scope of Computation}
When determining the scope of text on which to perform computation to determine which units of text to modify, we opted for paragraph-level over sentence- or document-level. This was motivated by our formative empirical observations when prototyping at each level.
When considering each sentence in isolation, relatively little text within the sentence was de-emphasized because we attempt to constraint the core meaning of the un-faded text to be very close to the original text---in this case, the sentence itself. 

However, many paragraphs have an overarching single topic, especially in certain kinds of writing like non-fiction, and using the paragraph as the scope of computation provides more leeway to de-emphasize parts of sentences (and sometimes, eventually even entire sentences) while still not straying too far from the overall core meaning of their containing paragraph, as captured computationally in \isac{} by LLMs. 
In other words, a typical paragraph is large enough to yield significant amounts of text for de-emphasis, but small enough to have a single coherent theme that is the focus of summarization.

Choosing something larger than paragraphs, such as entire documents, poses the challenge of the computational method making even larger choices about what to de-emphasize that a given reader might disagree with; in other words, it would be deciding on a larger, more noticeable scale which set of ideas are most critical to retain un-faded. \isac{}---no matter what scale of decisions an AI is making---allows readers to notice and, without taking any additional action, recover from differences of 'opinion' between the user and the AI, but \isac{} does not prevent annoyance. The larger the scale that the AI is ``getting it wrong'' (for the user), the more likely a reader may turn \isac{} off altogether.

\subsubsection{Extractive summarization method}
There are many methods of extractive summarization.
We started by exploring classical syntax-based methods, but found that parser errors and the limited flexibility of pruning parse trees led to output that was ungrammatical and/or missing key words.
Specifically, we tried running a dependency parser and shortening the sentence by removing subtrees based on depth and dependency type, similar to Filappova et al.~\cite{filippova2008dependency}.
This frequently removed important contextual information such as key adjectives and noun phrases.
Most existing extractive summarization methods also failed to achieve the desired results as they could not be used to generate our multi-level recursive extractive summary.
These observations led us to explore the potential of an LLM-based approach, given their recent improvements~\cite{tao2023eveval, riccardi2023two, hagendorff2023human}.

\subsection{Overview of the \isac{} System}

\begin{figure}
\centering
\includegraphics[width=0.7\columnwidth]{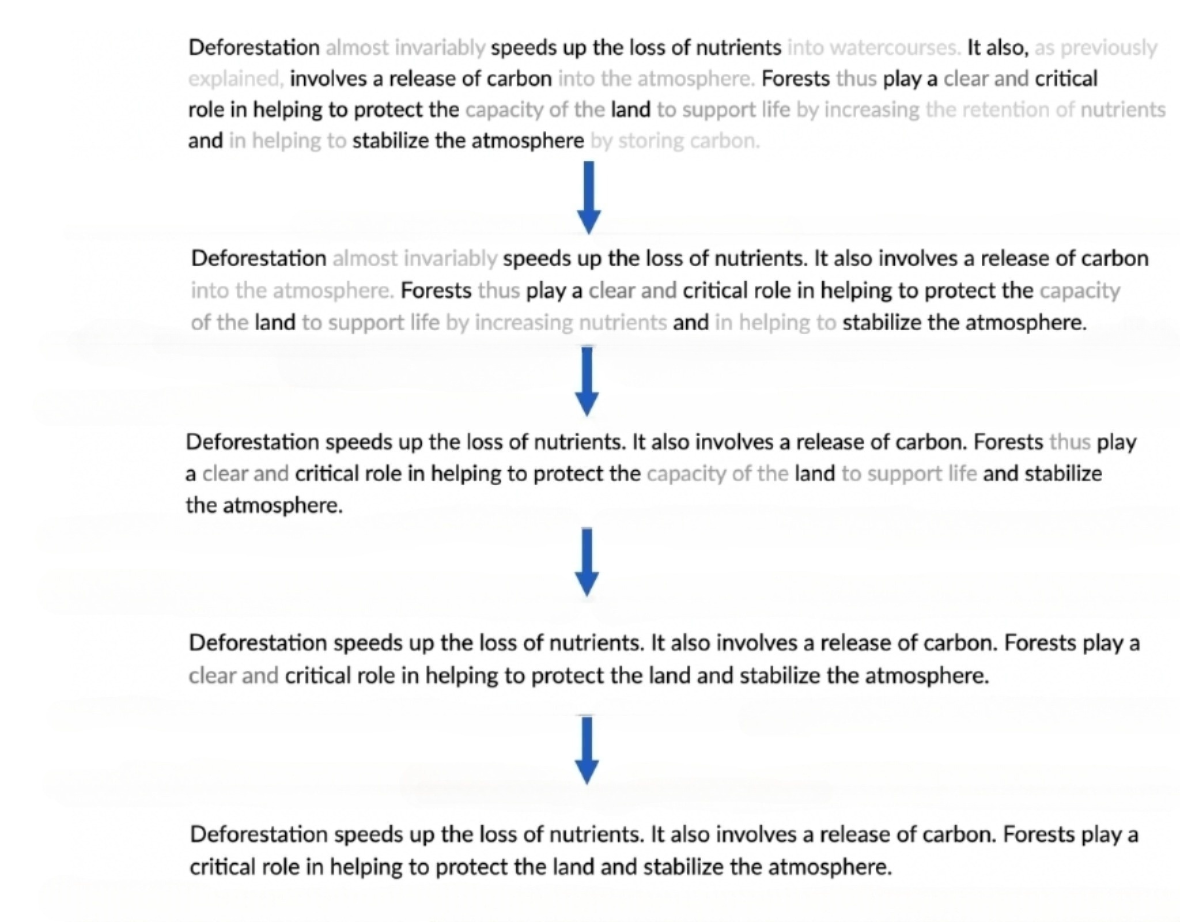}
\caption{An illustration of how a paragraph shortens with each round of extraction. Each level stays grammatical after shortening. The increasingly faded text at each level before the final most concise extractive summary show what will be removed at each level; the most faded text at the top level was removed first. What is rendered at the top level in this figure is the only rendering of this process that readers see.}
\label{fig-rounds}
\Description{Flowchart showing how a single paragraph gets successfully shortened by illustrating the five levels of granularity and the opacity of each word at each level. The first level is 'Deforestation almost invariably speeds up the loss of nutrients into watercourses. It also, as previously explained, involves a release of carbon into the atmosphere. Forests thus play a clear and critical
role in helping to protect the capacity of the land to support life by increasing the retention of nutrients and in helping to stabilize the atmosphere by storing carbon.' The second level is 'Deforestation almost invariably speeds up the loss of nutrients. It also involves a release of carbon into the atmosphere. Forests thus play a clear and critical role in helping to protect the capacity
of the land to support life by increasing nutrients and in helping to stabilize the atmosphere.' The third level is 'Deforestation speeds up the loss of nutrients. It also involves a release of carbon. Forests thus play
a clear and critical role in helping to protect the capacity of the land to support life and stabilize the atmosphere.' The fourth level is 'Deforestation speeds up the loss of nutrients. It also involves a release of carbon. Forests play a clear and critical role in helping to protect the land and stabilize the atmosphere.' The fifth (most succinct) level is 'Deforestation speeds up the loss of nutrients. It also involves a release of carbon. Forests play a
critical role in helping to protect the land and stabilize the atmosphere.'}
\end{figure}

The \isac{} visualization re-renders plain text at multiple levels of opacity; these levels reveal multiple successive recursive levels of grammatically correct detail within each sentence of each paragraph. It is produced by successively shortening the passage across multiple rounds (Figure~\ref{fig-rounds}). Words deleted in the first round are deemed the \emph{least} important, and therefore given the least opacity; words deleted in the second round are deemed slightly more meaningful and appear more opaque; and so on. Words that are never removed remain in full color. \textcolor{GRAY2}{In other words, }the \isac{} visualization operates like this: \textcolor{GRAY3}{some }text\textcolor{GRAY3}{,} not \textcolor{GRAY3}{entirely }relevant \textcolor{GRAY1}{to the core meaning }\textcolor{GRAY3}{of a sentence, }appears lighter than \textcolor{GRAY3}{relatively }\textcolor{GRAY1}{more }important text. When a sentence is \textcolor{GRAY3}{artificially }long \textcolor{GRAY3}{and complicated }\textcolor{GRAY2}{and full of irrelevant }\textcolor{GRAY3}{continuations and }\textcolor{GRAY2}{phrases }\textcolor{GRAY3}{that add little to the overall meaning, the }opacity \textcolor{GRAY1}{of }\textcolor{GRAY3}{certain }\textcolor{GRAY1}{words }\textcolor{GRAY3}{and phrases} is reduced based on \textcolor{GRAY3}{the outcome of }\textcolor{GRAY1}{successive }\textcolor{GRAY3}{rounds of }shortening. When a sentence is simple, words remain salient.\footnote{This fading of text, from ``In other words...'' onwards, was generated directly from our LLM-based method  described in Section~3.3 Implementation.}

\subsection{Algorithmic Workflow}
\label{sec:implementation}
Producing the \isac{} visualization for a given passage is nontrivial because it involves ensuring that every level of extraction remains both grammatical and sufficiently close to the core meaning of a passage, for some designer-set threshold and notion of closeness.
Our approach is powered by a large language model (LLM). Specifically, we prompt OpenAI's GPT4 with a single paragraph at a time and ask it to:\\

\noindent \emph{``Delete spans of words or phrases from the following paragraph that don't contribute much to its meaning, but keep readability:\newline\{paragraph\}\newline Please do not add any new words or change words, only delete words.''}\\

\noindent Though an LLM-based approach seemed fairly successful in our pilots and within the studies reported, we reflect on the inherent limitations of our choice of using an AI tool, especially its non-determinism, in the Discussion section.

While leveraging an LLM is the computational method behind our recursive sentence compression approach, simply asking an LLM to do this is insufficient on its own for three reasons: (1) sometimes it adds or changes words, (2) the quality of the output varies, and (3) it only provides one set of words to de-emphasize.
Our approach incorporates solutions to each of these:

\paragraph{Undoing LLM-inserted words and substitutions}
We use a SequenceMatcher\footnote{https://docs.python.org/3/library/difflib.html} to identify words that the LLM has added or changed.
These represent rewrites and are hence not allowed as they would mean the user is no longer seeing the original paragraph.
We replace substitutions with the original words from the paragraph, and remove insertions; the result is the \textit{post-reversion} LLM response.

\paragraph{Improving output quality}
Whenever the LLM generates a shortened paragraph, it may fall short of fulfilling its prompt, e.g., by removing words that leads to grammatical errors; only adding or substituting words; or removing words in a way that changes the meaning of the text too significantly.
We address this by always requesting the same LLM-shortened paragraph multiple times (i.e., $8$ times in our case) using the same prompt; empirically, we have observed that usually at least one resulting paragraph is sufficiently high quality for \isac{} to continue with.

To automatically identify the highest quality response, we composed a custom heuristic evaluator.
This heuristic evaluation assesses response quality based on a combination of four scores: semantic fidelity, response length, paraphrasing frequency, and grammatical correctness.
The semantic fidelity score is the similarity between the \textit{original} (pre-summarization) paragraph and the shortened paragraph, calculated using the cosine similarity of their respective embeddings produced by Sentence Transformers~\cite{reimers2019sentence}.
The length score measures how closely the response's length aligns with a preset optimal length, which, based on prototyping, was set to 85\% of the previous level's length. The paraphrasing metric quantifies the inverse of detected insertions and substitutions (as determined by a SequenceMatcher)---before such insertions and substitutions were automatically reverted. The grammaticality score involves re-prompting GPT4 to evaluate the syntax of the response after reversion, on a crude scale: 0 for `bad grammar', 0.5 for `moderately grammatical', and 1 for `grammatically correct'.\footnote{The prompt we used was: \emph{``Score the following paragraph by how grammatical it is.\newline\{paragraph\}\newline Answer A for grammatically correct, B for moderately grammatical, and C for bad grammar. Only respond with one letter.''} An A was mapped to a 1, a B was mapped to 0.5, and a C was mapped to 0.} All four scores are scaled to range from 0 to 1.
These scores are combined, via averaging, to produce an overall quality measure of each individual post-reversion LLM-shortened paragraph (one of the eight for every recursive round).
Finally, we select the highest scoring option, discard the rest, and proceed to the next level, which takes as input the highest-scoring LLM output that was just chosen. 

\paragraph{Identifying multiple levels of relevance}
For each paragraph in the given passage, we run multiple rounds of LLM-powered extractive paragraph summaries---each on the results of the previous round---to identify multiple levels of criticality within each sentence of each paragraph.
In each round, we use the methods described above to (a) request 8 responses from GPT4, (b) resolve word addition and substitution, and (c) select the best option using the evaluator.
In the first round, the input is the entire paragraph.
In subsequent rounds, the input is the best output from the previous round.

This recursive extractive summarization process stops when the LLM ``refuses'' to cut any words from the summary chosen for the ``deepest'' level reached so far. We chose this stopping criterion after observing that the LLM will often return the paragraph unchanged if it cannot find additional words to delete, and that this is a better stopping criterion than any other heuristic we experimented with because it is sensitive to the complexity of the original paragraph. More complex paragraphs can accommodate more recursive levels of summarization, while simpler paragraphs may have very few words that can be cut and still maintain grammaticality. So our recursive process stops when none of the 8 requested LLM summarized paragraphs return a summary with any deleted words.

\subsubsection{Implementation}
The interface is a Flask application that is configured to make API calls to OpenAI's GPT4 model. The web application is publicly available at [REMOVED FOR REVIEW]. Our code is publicly available at [REMOVED FOR REVIEW].

\section{User Studies}

We evaluated \isac{} in two studies---a preliminary user study of the effectiveness of the visualization given a partially automated backend and a summative user study that measures the impact of the \isac{} when fully automated. In every user study, every interface being tested was referred to by an arbitrarily assigned color, e.g., "reader-green" or "reader-blue", as others have done previously, e.g.,~\cite{singh2022hide}. 

\subsection{Preliminary User Study}
\subsubsection{Overview}

To understand whether the \isac{} visualization we propose improves reading comprehension and the reading experience, we first conducted a preliminary user study with 18 participants involving a semi-automated human-supervised version of \isac{}. In this phase of our work, our aim was to gauge the efficacy of modulating text opacity over nested grammatical subsets of sentences while setting aside concerns about the quality of the backend. In other words, we wanted to verify that grammar-preserving text saliency modulation actually helps, if the eventual fully-automated AI backend is able to perform as well as the human-in-the-loop (partially automated) AI backend we used in this study. 

Our decision to employ such a partially automated approach stems from emerging practices in prototyping AI and NLP systems~\cite{yang2019sketching, yang2023harnessing}, which argue that, given the significant effort and time required to verify output quality of a production-ready AI-powered system, Wizard of Oz-like techniques that employ human-verified AI outputs should be used first before deciding whether to implement the actual AI system.

Our preliminary study evaluates the exact same visualization as the eventual fully automated \isac{} system, but instead of automatically choosing the best response from GPT4, a human inspector picked the response they believed was best, e.g., had the least rephrasing, and then manually reverted any rephrases in the chosen response. When necessary, the human inspector also edited the response to fix ungrammaticality. 

One additional interactive variant was included as an extra condition for comparison in this preliminary study, which was not included in the final fully automated \isac{} evaluated in the second study. In this variant, a slider or mousewheel affordance can be used to hide text below a certain level of opacity. It, however, partially violates the design goals: even though it is trivial to reveal hidden levels of detail by moving the slider, unless the slider to set to its lowest setting, which is equivalent to the static (and final) version of \isac{}, it is not possible for the reader to notice and recover from automated decisions they disagree, unless they remember what was hidden.

We were interested primarily in the following questions:

\begin{itemize}
  \item \textit{How does the \isac{} visualization affect people in reading and skimming?}
  \item \textit{What is the user experience like when using the \isac{} visualization for reading and skimming?}
  \item \textit{What kinds of value, if any, does interactive granularity control provide for readers?}

\end{itemize}

\begin{figure}[h]
    \includegraphics[width=0.9\textwidth]{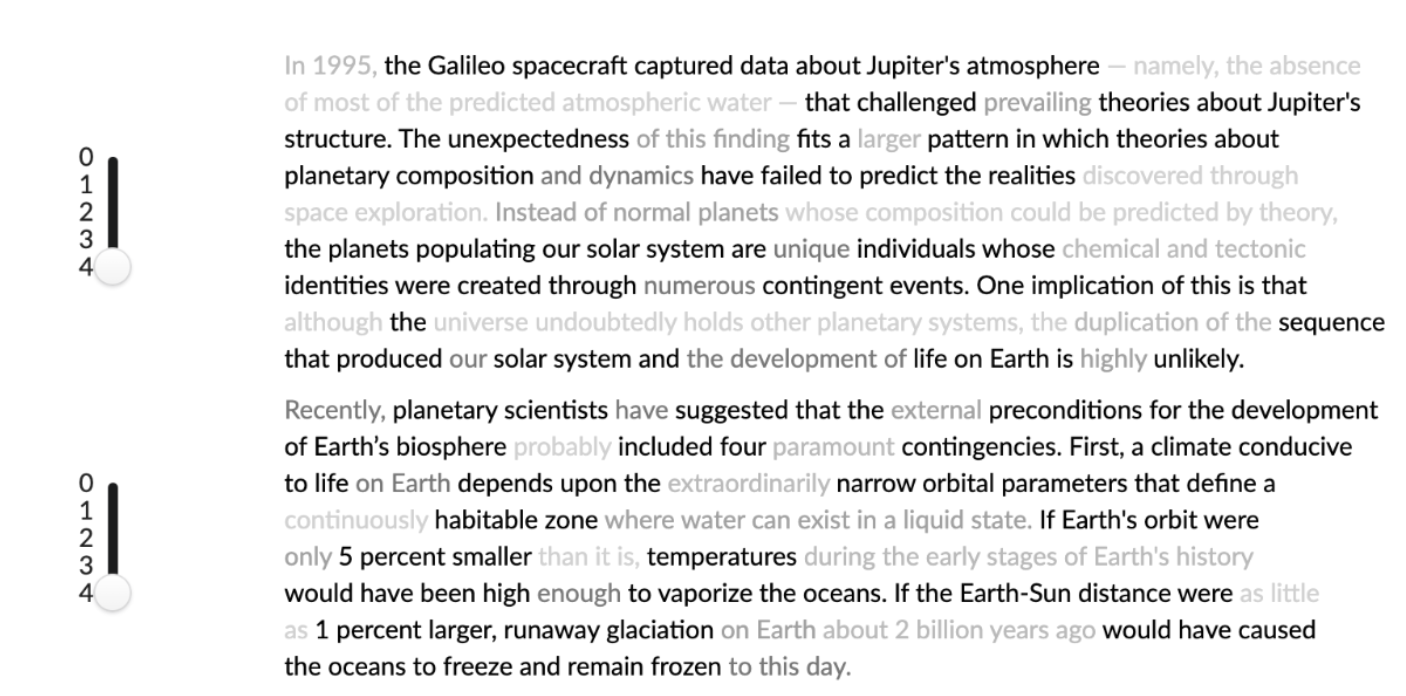}
    \caption{Screenshot of the \interactive{} interface in the preliminary user study. The (static) \static{} interface is exactly the same, but without the sliders or the responsiveness to mouse scrolling to hide segments of text below a certain level of opacity.}
    \label{fig:prelim_study_interfaces}
\end{figure}

In summary, to study these questions, we designed a within-subjects design with three conditions: \static{} (Human-in-the-Loop GP-TSM), \interactive{} (Human-in-the-Loop GP-TSM with interactive granularity), and \control{}. \static{} is our partially-automated \isac{} visualization, with only a simple toggle to turn it on and off; \interactive{} added interactive granularity to \static{}; and \control{} was simply presenting the original plain text. All conditions used the same font and font size (Lato, 14pt). Figure~\ref{fig:prelim_study_interfaces} presents a screenshot of the \interactive{} condition.

\subsubsection{Procedure, Participants, Conditions, and Measures}
\label{sec:procedure}

We recruited 18 participants (8 female and 10 male; 8 between 19-24 years of age and 10 between 25-34 years of age) from university mailing lists at an R1 university in North America. Participants received a \$20 Amazon gift card as compensation. Our screen criteria was: \emph{``Participants need to be fluent in English and over 18 years of age.''} Participants' self-reported English reading proficiency was relatively high (asked to rate proficiency out of ten, with ten highest: M=8.38 (SD=1.37). 

Our study was split into the following parts: informed consent, three sequential reading tasks, and a final survey. Prior to starting each task, each participant went through a short walk-through of the task and affordances of the assigned interface condition. Each participant was given up to 10 minutes to complete each reading task, and asked to complete the task as fast as they could to the best of their ability. The entire study took about 60 minutes. 

Each reading task was completed in a separate interface condition ({\static{}}, {\interactive{}}, or {\control{}}). Participants encountered each interface and each reading task exactly once, and both the reading task order and condition order was counterbalanced across participants. Specifically, we performed a partial counterbalancing of passages to conditions that ensured each passage appeared the same number of times in each condition, and in each condition in each position. Were there any substantial differences in difficulty between passages, this counterbalancing reduces the effect such a difference may have, however we only sampled half of an entire counterbalancing set, which is why subsequent analysis described in the results uses a mixed effects model. We refer to passages as R1, R2, and R3.

We chose Graduate Record Examinations (GRE) passages and reading comprehension questions\footnote{All the passages and questions we used are from publicly available GRE Practice Tests provided by the Educational Testing Service (ETS).} as our tasks, specifically the `Long Passages' subsections of the GRE Verbal Reasoning section, each with exactly four questions. They are a relatively standardized measure of reading comprehension; they are specifically designed to require close reading, measure participants' understanding of the text, are standardized to have similar difficulty, and all questions count equally towards the final score~\cite{gre_reading_tests}. Notably, the three selected passages are of comparable length, with word counts of 472, 446, and 444, respectively. 

After each reading task, participants completed a questionnaire to record their reflections on their experience and perceived difficulty of the task in the assigned condition. Questions included an overall rating of the interface and NASA TLX survey questions, and two questions about self-rated task performance.
The \interactive{} and \static{} conditions had four additional questions about the visualization; and \interactive{} had another two additional questions about the interactive granularity. After finishing the reading tasks, participants were asked to fill out a post-study survey to indicate their preferences across all three conditions and provide further qualitative feedback. Post-study surveys are provided in Appendix~\ref{appendix:survey}.

\subsubsection{Results}
\label{sec:prelimresults}

\newcommand{\anovaprob}[4]{($p{=}{#1}$, $F_{{#2},{#3}}{=}{#4}$)}
\newcommand{\meansd}[2]{(M={#1}, SD={#2})}

We analyzed reading task results with a three-factor (repeated measures) ANOVA mixed effects model; specifically, investigating each dependent variable on fixed factors \emph{Condition}, \emph{Passage}, and \emph{Order} (the position of the task, first second or third in the sequence) and any interaction effects among these factors, controlling for the random factor of \emph{Participant}. Satterthwaite's method was used to estimate denominator degrees of freedom. Pairwise comparison with Tukey's HSD (with $\alpha{=}0.05$) was conducted between each of the three conditions and three passages. Hereafter, we refer to these methods as ANOVA and Tukey's test, respectively. 

\begin{figure}[h]
    \centering
    \begin{minipage}[b]{0.45\textwidth}
        \includegraphics[width=\textwidth]{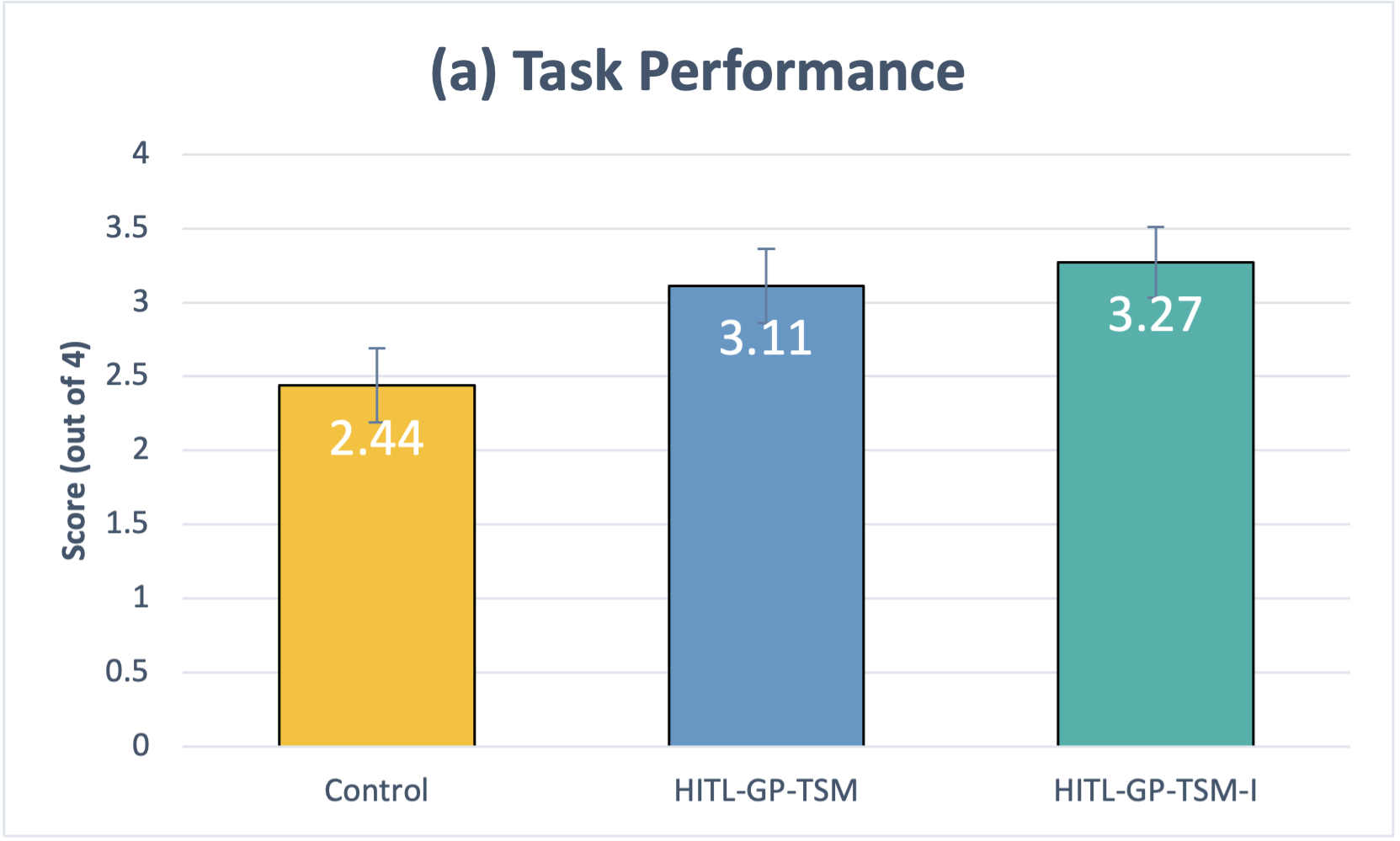}
    \end{minipage}
    \hfill
    \begin{minipage}[b]{0.45\textwidth}
        \includegraphics[width=\textwidth]{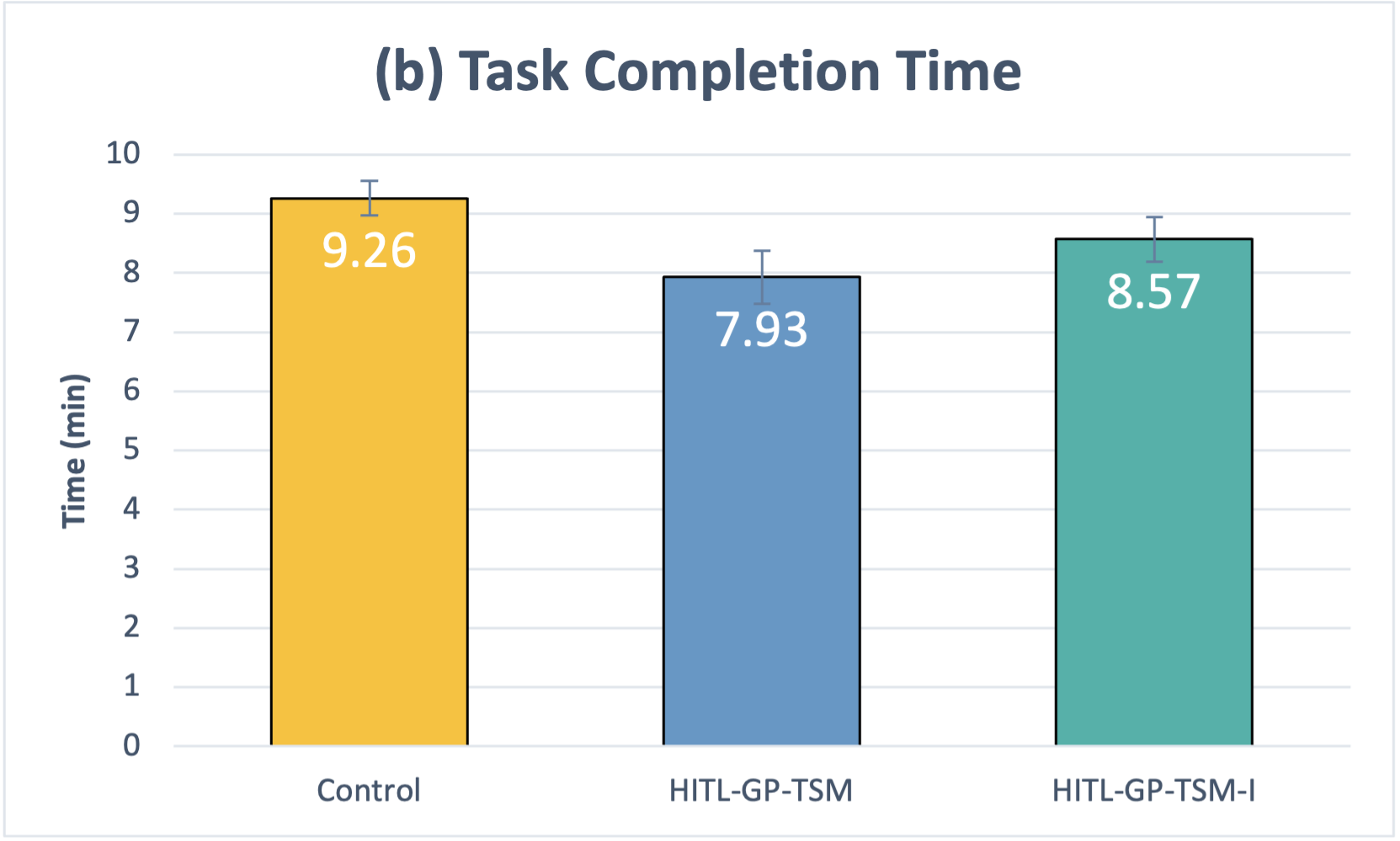}
    \end{minipage}
    \caption{In the preliminary user study, \interactive{} resulted in significantly better performance on the reading comprehension task than \control{}---on the order of nearly an entire reading comprehension question out of a total of 4, though participants in the \static{} condition were not far behind. In the \static{} condition, participants completed their reading comprehension tasks significantly faster than when using the \control{}. The error bars represent standard error.}
    \label{fig:prelim_study}
\end{figure}

ANOVA analysis shows a significant main effect of Condition on reading comprehension scores ($p{=}.03$, $F_{2,11.3}{=}4.81$). Using Tukey's test, we found that, compared to participants using \control{}, participants with access to interactive granularity (\interactive{}) scored significantly better on reading comprehension questions($p{=}0.047$)---answering approximately three fourths of an additional question correctly out of four. Participants in \static{} were not far behind, though the difference was not significant---they answered approximately half an additional answer correctly out of four, relative to participants in \control{}. This can be seen in Figure~\ref{fig:prelim_study}.

ANOVA analysis also found a significant main effect of Condition on time spent completing each reading task ($p{=}.022$, $F_{2,10.6}{=}5.52$). Participants using \static{} completed their reading comprehension questions in only 7.9min (SD=1.9min) on average, which Tukey's test shows was significantly faster ($p{=}0.029$) than participants in \control{}, which completed their reading comprehension questions, on average, 1.4 minutes later, at 9.3min (SD=1.2min). Other tests do not reach significance. 

Participants using \interactive{} were not significantly faster than \control{}, but this may have been due to effects that would ultimately fade with additional use if this were deployed in the wild, i.e., some participants spent some of their time playing with the interactive elements, ``trying out different widgets'' and ``figuring out what exactly the mouse and sliders do.'' 

\begin{table}
\small
    \begin{tabular}{l|l|l|l}
        \toprule 
        \textbf{Question Statements}  & \textbf{GP-TSM} & \textbf{GP-TSM-I} & \textbf{Control}  \\ \midrule 
        How would you rate your overall experience in this interface?                                      & 5.61 (1.24)*       & 5.44 (1.25)*                   & 4.06 (1.21)      \\ \midrule
        How mentally demanding was the task? \textit{\textbf{[Lower is better (LIB)]}}                                                              & 4.33 (1.37)*       & 4.56 (1.46)                    & 5.5 (1.29)       \\ \midrule
        How physically demanding was the task? \textit{(LIB)}                                                            & 1.94 (1.35)        & 2.28 (1.41)                    & 2.67 (1.71)      \\ \midrule
        How hurried or rushed was the pace of the task? \textit{(LIB)}                                                    & 2.72 (1.45)*       & 3.17 (1.62)                    & 4.28 (1.67)      \\ \midrule
        How successful do you think you were in accomplishing the task?                                    & 5.11 (1.41)        & 5.33 (1.14)                    & 4.56 (1.50)      \\ \midrule
        How hard did you have to work to accomplish your level of performance? \textit{(LIB)}                            & 3.94 (1.21)        & 4.17 (1.20)                    & 4.94 (1.43)      \\ \midrule
        How insecure, discouraged, irritated, stressed, and annoyed were you during the task? \textit{(LIB)} & 2.67 (1.50)        & 3.06 (1.55)                    & 3.78 (1.90)      \\ \midrule
        I could recognize the key points in the passage.                                                   & 6.11 (1.02)*       & 5.89 (1.08)                    & 5.0 (1.37)       \\ \midrule
        I could recognize how the key points are supported by additional detail in the passage.            & 5.89 (1.13)*       & 5.5 (1.29)                     & 4.78 (1.56)      \\ \midrule
        The system’s choice of what to gray out and what to keep at full font weight made sense to me.     & 5.61 (1.42)        & 5.78 (1.31)                    & N/A              \\ \midrule
        I think I know why certain words were lighter than others.                                         & 5.67 (1.24)        & 5.83 (1.25)                    & N/A              \\ \midrule
        I found it helpful that certain words were lighter than others.                                    & 5.44 (1.20)        & 5.67 (1.28)                    & N/A              \\ \midrule
        The different levels of gray helped me see the relationships between different parts of sentences. & 5.28 (1.23)        & 5.22 (1.35)                    & N/A              \\ 
        \bottomrule 
    \end{tabular}
    \caption{Statistics of scores in the survey after each reading task. For brevity, we use \isac{} for \static{} and \isac{}-I for \interactive{}. Participants were asked to rate their agreement with statements related to their reading experience on a 7-point Likert scale from ``Strongly Disagree'' (1) to ``Strongly Agree'' (7). The questions 2 through 7 (and their scales) were adapted from the NASA Task Load Index~\cite{hart2006nasa}. \textit{``LIB''} stands for \textit{``Lower is better.''} Statistics in column 2, 3, and 4 are presented in the form of mean (standard deviation). ANOVA analysis shows a significant main effect of Condition on participants' answers. Statistically significant (p < 0.05) differences compared with \control{} through Tukey's HSD tests are marked with a *. For the last four statements, which concern the text opacity visualization and thus do not apply to the control condition, significance was calculated based on just the two remaining experimental conditions.} 
    \label{tab:prelim_study}
\end{table}

Participants generally expressed a preference for the 2 \isac{} conditions over \control{}. 
ANOVA analysis shows a significant main effect of Condition on participants' answers to the questions in Table~\ref{tab:prelim_study}, which were asked after each reading task. Tukey's test shows that \static{} received significantly more positive ratings from participants than \control{} in 5 out of the 9 questions that were asked in both conditions, for \textit{overall experience, how mentally demanding the task was in that condition, how hurried or rushed they felt,}, \textit{ability to recognize key points in the passage} and \textit{ability to recognize how key points were supported by additional detail}. \interactive{} only received significantly more positive ratings from participants compared to \control{} on the question about \textit{overall experience}. 

These preliminary results verify the usability and helpfulness of the \isac{} visualization in supporting reading comprehension, suggesting that it would be worthwhile to implement and evaluate a fully automated version of \isac{}. 

The benefits of interactive granularity were less clear cut. While the \interactive{} condition also results in significantly better performances and reading experience than \control{}, the difference between \interactive{} and \static{} is not significant. Moreover, only \static{} results in both significantly faster task completion and significantly lower perceived difficulty. Therefore, we decided not to carry the feature of interactive granularity forward into the next stage of development.

\subsection{Main User Study of the fully automated \isac{}}
\label{sec:mainstudy}

\subsubsection{Overview}
After implementing a fully automated version of \isac{}, described in Section~\ref{sec:implementation}, we conducted a user study with a separate set of 18 participants to evaluate the efficacy of the fully automated static \isac{}, using a very similar study format. This time, we were interested primarily in the following questions:

\begin{itemize}
  \item \textit{How does the fully automated \isac{} affect reading comprehension? Reading experience?}
  \item \textit{How does \isac{} compare to the nearest previously published text saliency modulation method for reading and skimming?}
  \item \textit{Is the rendering of multiple nested grammatical subsets of sentences resulting from recursive extractive summarization intelligible to users?}
  \item \textit{What is the impact of preserving the grammaticality of each nested subset of each sentence in \isac{} provide on users, relative to a version of \isac{} that does not preserve grammaticality?}
\end{itemize}

\noindent To answer these questions, we modified the preliminary study design in the following ways: 

First, the interactive granularity condition was replaced with \wf{}, which we identified as the nearest previously published text saliency modulation method for reading and skimming. As in~\cite{10.1145/1753846.1754093}, \wf{} modulates font opacity, but is based on unigram frequency~\cite{brath2014using, 10.1145/1753846.1754093}. In other words, words that appear less frequently are rendered more opaque in \wf{}, and more frequent words are less opaque. It is worth noting that the percentage of words that are less than fully opaque in \wf{} is comparable to that in the \isac{} condition, so any effects we observe are not due to how many words are grayed out, but \emph{which} words are grayed out.

Second, we added a second study component to the end, in which users experience and are asked to reflect on reading the same passage in two different conditions: \isac{} and a new control, \ngp{}. \ngp{} is \isac{} with grammaticality constraints removed from both places within its workflow: the LLM prompt  and the LLM response evaluator. Specifically, the phrase \emph{but keep
readability} in the \isac{} prompt was replaced with \emph{Don't worry about grammar},\footnote{The modified, non-grammar preserving extractive summarization prompt, in its entirety, was: \emph{``Delete spans of words or phrases from the following paragraph that don't contribute much to its meaning. Don't worry about grammar:\newline\{paragraph\}\newline Please do not add any new words or change words, only delete words.''}} and the grammaticality score was removed from the evaluation heuristic, and hence un-enforced. In other words, \isac{} enforces grammaticality at every minimum level of opacity and \ngp{} does not; asking participants to compare them helps us answer our research question about the criticality of grammaticality enforcment to the success of \isac{}.

Specifically, this means that after participants finished all the reading tasks and the post-all-reading-tasks survey that were present in both the preliminary user study and this user study, we asked them to participate in a 5-minute survey where we presented them with a view of the same passage rendered twice, side by side, once with \isac{} and once with \ngp{}. An example is included in Appendix~\ref{appendix:comparison}. We counterbalanced the presentation order of the two passages to ensure that each appeared on the left and right sides an equal number of times. We then inquired if participants could discern any differences between the two and, if so, to specify those differences. Additionally, we sought their preference between the two visualizations. This part of the study was exploratory and preliminary, meant to give us an indication of the value grammar preservation adds to our system.

\begin{figure}[h]
    \centering
    \begin{minipage}[b]{0.49\textwidth}
        \includegraphics[width=\textwidth]{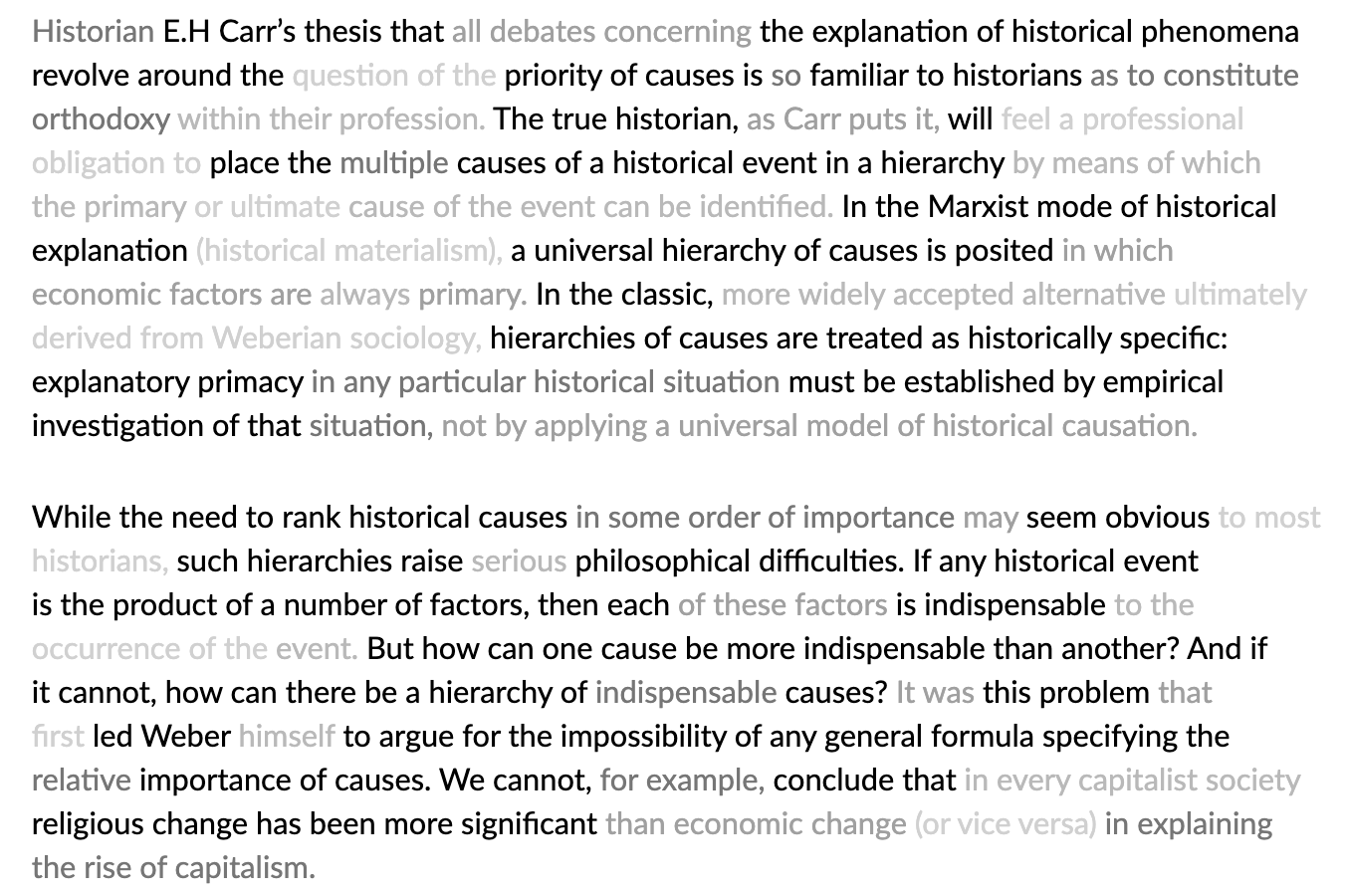}
    \end{minipage}
    \hfill
    \begin{minipage}[b]{0.49\textwidth}
        \includegraphics[width=\textwidth]{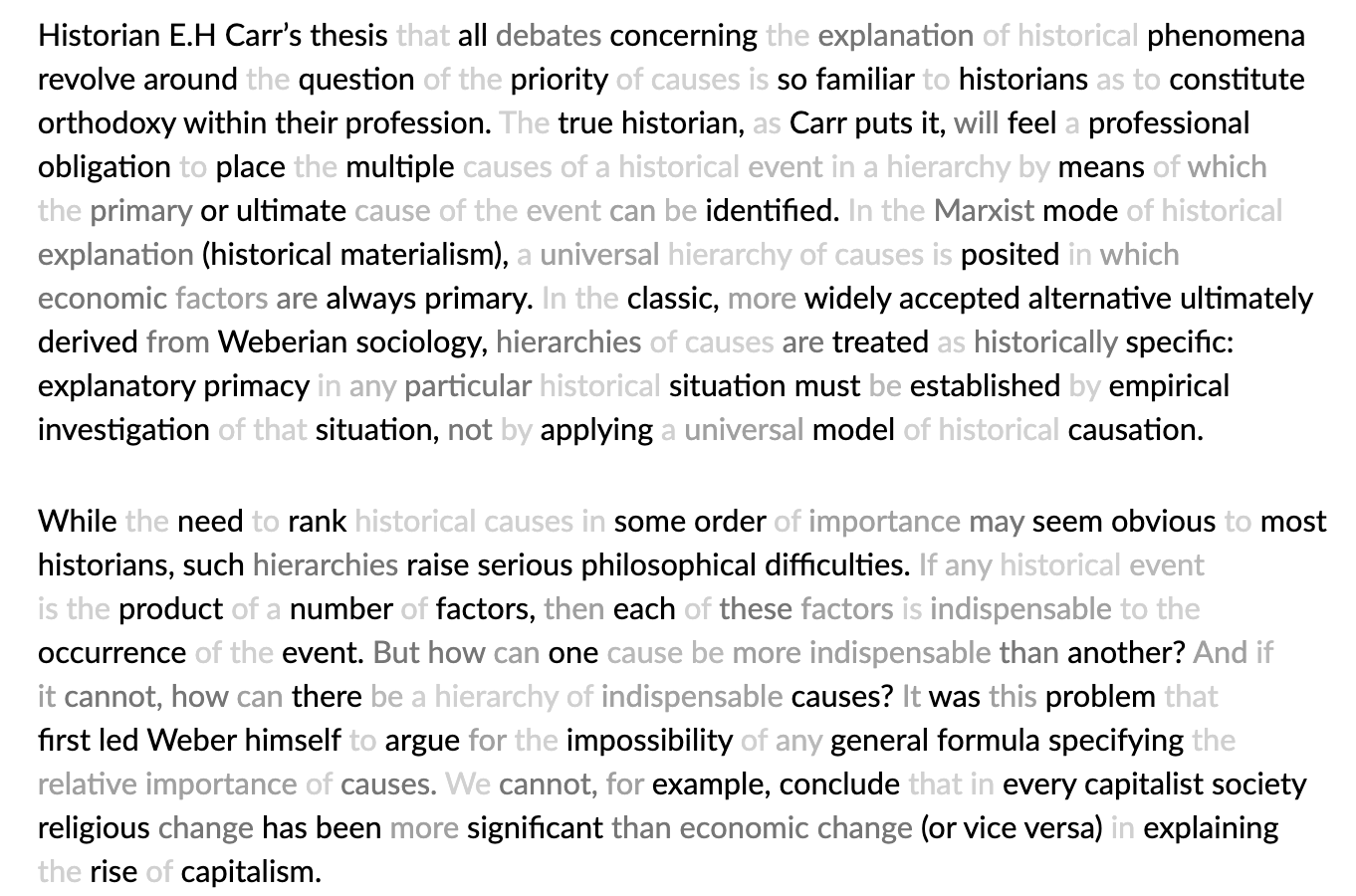}
    \end{minipage}
    \caption{Screenshots of the \isac{} (left) and \wf{} (right) interfaces in the main user study.}
    \label{fig:main_study_interfaces}
\end{figure}

All other aspects, including the \control{} condition, the chosen reading passages and the counterbalancing of conditions, passages, and their respective pairings, remained consistent with the preliminary study. Figure~\ref{fig:main_study_interfaces} presents screenshots of the \isac{} and \wf{} conditions in the main user study, each displaying the same passage.

While exact timing information was not recorded, the fully automated \isac{} took approximately 2-3 minutes to compute and render the GRE Long Passages texts for each reading task. This time did not affect participants' task time because the results were cached.

\subsubsection{Participants}
We recruited a separate set of 18 participants (7 self-identified as female, 10 as male, and 1 as non-binary; 7 were between 18-24 years of age, 10 between 25-34 years of age, and 1 between 35-44 years of age) from university mailing lists at an R1 university in North America. None of them previously participated in the preliminary user study. Participants received a \$25 Amazon gift card as compensation. Participants' self-reported English reading proficiency was relatively high, when asked to rate proficiency out of ten, with ten highest: M=8.59 (SD=1.71). 

\subsubsection{Quantitative Results}
We used the same statistical analysis process as in the preliminary study (Sec.~\ref{sec:prelimresults}) to analyze the reading task results and Likert survey questions. 

\begin{figure}[h]
    \centering
    \begin{minipage}[b]{0.45\textwidth}
        \includegraphics[width=\textwidth]{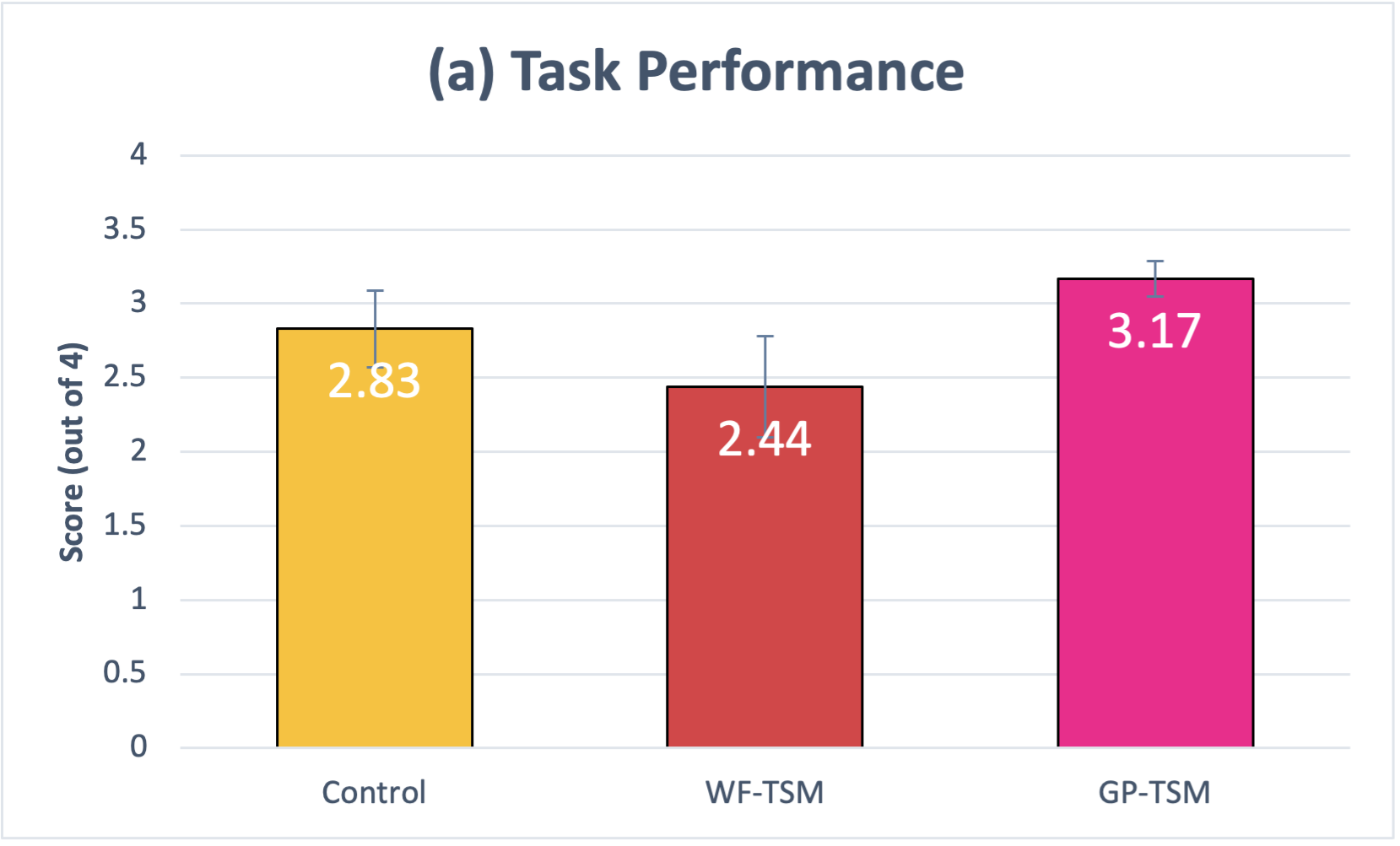}
    \end{minipage}
    \hfill
    \begin{minipage}[b]{0.45\textwidth}
        \includegraphics[width=\textwidth]{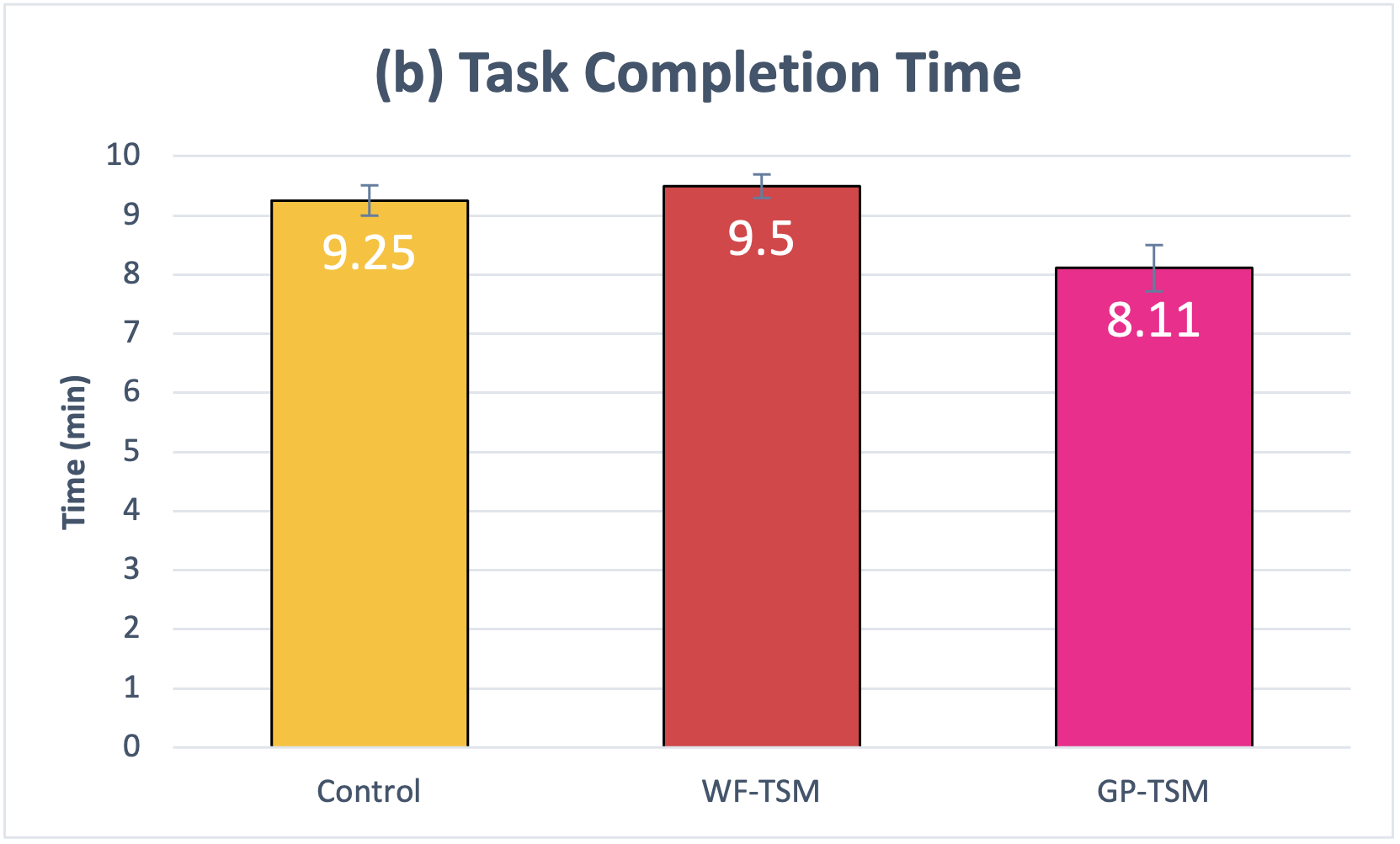}
    \end{minipage}
    \caption{Participants performed significantly better and significantly faster in the reading comprehension task when using \isac{} compared to the control conditions in the user study. The error bars represent standard error.}
    \label{fig:main_study}
\end{figure}

Overall, we find that participants performed significantly better when using \isac{} compared to \control{}, and when using \control{} compared to \wf{}. Participants also completed tasks significantly faster when using \isac{}, compared to both \wf{} and \control{} (Figure~\ref{fig:main_study}). Specifically, ANOVA analysis shows a significant main effect of Condition on reading comprehension scores ($p{=}.02$, $F_{2,11.3}{=}4.79$), and Tukey's test shows that participants earned significantly higher scores when using \isac{} compared to participants using \control{} ($p{=}.021$). Tukey's test also shows that \wf{} was significantly worse than \control{} ($p{=}.045$).

ANOVA analysis shows a significant main effect of Condition on task times while achieving these reading comprehension scores ($p{=}.0026$, $F_{2,10.6}{=}6.72$).
Tukey's test shows that, when using \isac{}, it took participants, on average 8min 7s to complete the reading task (SD=1min 36s), which was, on average, about a minute faster than when using the \control{} (M=9.25min, SD=1min 4s)($p{=}0.019$). Participants using \wf{} were slightly slower than the \control{} by 15 seconds on average (M=9.5min, SD=49s), though that difference was not significant. The difference between \isac{} and \wf, however, was still significant ($p{=}0.003$). 

\begin{table}
\small
    \begin{tabular}{l|l|l|l}
        \toprule 
        \textbf{Question Statements}  & \textbf{GP-TSM} & \textbf{WF-TSM} & \textbf{Control}  \\ \midrule 
        How would you rate your overall experience in this interface?                                      & 5.52 (1.45)*\dag       & 4.1 (1.19)                   & 4.35 (1.32)      \\ \midrule
        How mentally demanding was the task? \textit{\textbf{[Lower is better (LIB)]}}                                                               & 3.98 (1.46)*\dag       & 5.21 (1.48)                    & 5.13 (1.3)       \\ \midrule
        How physically demanding was the task? \textit{(LIB)}                                                            & 1.91 (1.34)        & 1.98 (1.54)                    & 1.85 (1.62)      \\ \midrule
        How hurried or rushed was the pace of the task? \textit{(LIB)}                                                   & 3.15 (1.88)       & 4.49 (1.91)                    & 4.35 (1.72)      \\ \midrule
        How successful do you think you were in accomplishing the task?                                    & 5.09 (1.37)*        & 3.87 (1.56)                    & 4.61 (1.48)      \\ \midrule
        How hard did you have to work to accomplish your level of performance? \textit{(LIB)}                            & 3.65 (1.16)*\dag       & 4.97 (1.52)                    & 4.78 (1.43)      \\ \midrule
        How insecure, discouraged, irritated, stressed, and annoyed were you during the task? \textit{(LIB)}  & 2.32 (1.38)*        & 4.57 (2.1)                    & 2.29 (1.31)      \\ \midrule
        I could recognize the key points in the passage.                                                   & 6.16 (1.04)*\dag       & 5.12 (1.11)                    & 5.23 (1.06)       \\ \midrule
        I could recognize how the key points are supported by additional detail in the passage.            & 5.9 (1.09)       & 4.98 (1.25)                     & 5.38 (1.46)      \\ \midrule
        The system’s choice of what to gray out and what to keep at full font weight made sense to me.     & 5.91 (1.67)*        & 2.38 (1.44)                    & N/A              \\ \midrule
        I think I know why certain words were lighter than others.                                         & 5.36 (1.14)*        & 4.83 (1.21)                    & N/A              \\ \midrule
        I found it helpful that certain words were lighter than others.                                    & 5.67 (1.81)*        & 3.12 (1.33)                    & N/A              \\ \midrule
        The different levels of gray helped me see the relationships between different parts of sentences. & 5.19 (1.22)*        & 2.41 (1.25)                    & N/A              \\ 
        \bottomrule 
    \end{tabular}
    \caption{Statistics of scores in the survey after each reading task. Participants were asked to rate their agreement with statements related to their reading experience on a 7-point Likert scale from ``Strongly Disagree'' (1) to ``Strongly Agree'' (7). Questions 2 through 7 (and their scales) were adapted from the NASA Task Load Index~\cite{hart2006nasa}. \textit{``LIB''} stands for \textit{``Lower is better.''} Statistics in column 2, 3, and 4 are presented in the form of mean (standard deviation). ANOVA analysis shows a significant main effect of Condition on participants' answers. Statistically significant (p < 0.05) differences compared with \wf{} and \control{} are marked with * and \dag, respectively. For the last four statements, which concern the text opacity visualization and thus do not apply to the control condition, significance was calculated based on the two experimental conditions.} 
    \label{tab:main_study}
\end{table}
Table~\ref{tab:main_study} shows participants answers to questions asked immediately after each reading task, with some exceptions when the questions are irrelevant in a given condition. ANOVA analysis shows a significant main effect of Condition on their answers. 

Overall, participants generally expressed preference for \isac{} over \wf{} and \control{}. Tukey's test shows that \isac{} received significantly more positive ratings from participants than \control{} in 4 out of the 9 questions that were asked in both conditions ($p{<}0.05$), for \textit{overall experience, how mentally demanding the task was in that condition, how hard they had to work in that condition,} and \textit{recognizing key points in the passage}. \isac{} also received significantly more positive ratings from participants than \wf{} in 6 out of the 9 questions that were asked in both conditions ($p{<}0.05$). These questions included all of the same questions that were significant for the \isac{}-\control{} comparison and additionally included \textit{how successful they thought they were} and \textit{how insecure, discouraged, etc. they felt}. 
Finally, in the questions which were only asked in the \isac{} and \wf{} conditions because they asked specifically about opacity modulation which was not present in \control{}, \isac{} received significantly better Likert scale ratings than \wf{} for all 4 questions (last 4 rows of Table~\ref{tab:main_study}), which were all about the \textit{intelligibility of why certain words were less salient} and their \textit{helpfulness}, especially for \textit{seeing the relationships between different parts within sentences}.

After experiencing all the conditions, participants were asked to rate their agreement on a 7-point scale (7 being the highest) with the following statement for each condition: \emph{``I would like to use [Condition] to read online text of interest to me in the future''}. \isac{} received a mean agreement score of 5.8 (SD=1.8) while \control{} received a mean agreement score of 4.7 (SD=1.3) and \wf{} received a mean agreement score of 3.3 (SD=2). This difference was significant ($p{<}0.05$) between \isac{} and both \control{} and \wf{} using additional pairwise unpaired t-tests. 

\begin{figure}[h!]
    \centering
    \includegraphics[width=0.65\textwidth]{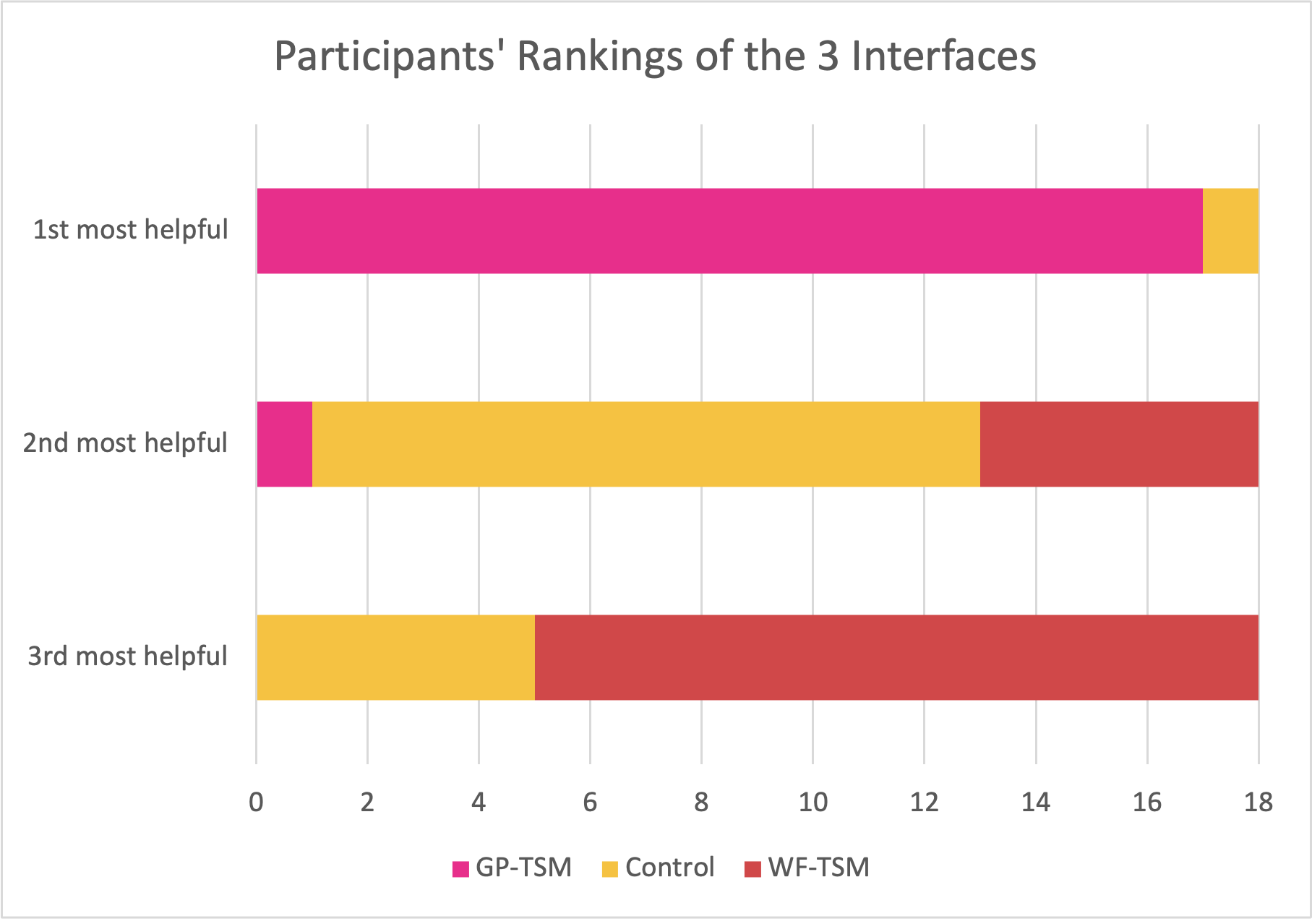}
    \caption{Participants' rankings of the 3 conditions in terms of their helpfulness for reading comprehension tasks}
    \label{fig:ranking}
\end{figure}

After experiencing all the conditions, participants were also asked to directly rank conditions. Participants expressed a strong preference for \isac{} over \wf{} and \control{}  (Figure~\ref{fig:ranking}). 

\subsubsection{Qualitative Feedback}
Overall, participants were positive about \isac{} and its functionality. Below, we group participants' responses to the survey questions\footnote{What did you like about the interface? \newline What did you not like about the interface? \newline What do you wish the interface had?} around a set of themes that were frequently mentioned. 

\paragraph{Improved Reading Efficiency} Twelve out of 18 participants appreciated how text saliency was modulated on the \isac{} interface, noting that it facilitated more efficient reading by letting them skip words but still grasp the gist of the passage (P1, P2, P3, P5, P6, P7, P9, P12, P14, P15, P16, P18). For example, P2 wrote, \emph{``My reading became faster and less congested [using \isac{}] because I could easily skip over and ignore words that were grayed out if I simply wanted to get the main idea of the passage.''} Further, P1 commented on how easy it was for \emph{``the structural logic of the passage to be absorbed''} given \isac{}'s multi-level visualization.

As for the mechanism by which reading efficiency was improved, participants provided interesting reflections on how \isac{} shaped their reading experience. It seemed that \isac{} facilitated a two-step reading process where participants \emph{``first skimmed the black words for a quick grasp of the gist and then went back to certain sections to read more details''} (P17). For instance, P6 observed, \emph{``The interface almost allowed me to read the passage in two ways. One way was to read each word, regardless of color like normal. But, alternatively, if I just read the black text I got the crux of the argument with none of the additional filler.''} P10 reflected on their question answering procedure specifically: \emph{``I was better able to search back through the text to find key words or ideas that related to the questions I was trying to find answers for. I felt that I could continuously read the bold words and they formed understandable sentences.''}

\paragraph{Sensible Visualization} Participants highlighted that the graying was \emph{``pretty consistent and reasonable''} to them (P9), with a lot of comments on how they agreed with what the system chose to gray (P2, P7, P12, P13, P16, P17), though P14 complained about the fact that sometimes \emph{``certain transition words are in gray,''} which \emph{``prevents [P14] from picking up the transition logic between two elements in a sentence, or between two sentences.''}

Participants also commented on the text visualization itself, pointing out that the graying was \emph{``natural''} (P11) and \emph{``not too much drama but provided just the right amount of contrast for engaging reading''} (P14). Overall, participants found that the system's design helped them focus on key points and read more efficiently (P1, P3, P7, P8, P16, P17). Some mentioned they were excited about using \isac{} in the future (P11: \emph{``I would want to use [\isac{}] to read my history readings.''}), possibly as a Chrome extension.

\paragraph{Explanation Needed} Despite a broad preference for the \isac{} interface, participants also offered several suggestions for improvement. A recurrent theme was the desire for more explicit guidance, which reveals that not all participants quickly and intuitively grokked what the levels of gray meant. For example, P2 suggested that  \emph{``a prompt that explained why some words were grayed out could have been helpful.''} Similarly, P12 noted that \emph{``it would have helped to have a tutorial to understand why some text was more gray.''}

\paragraph{Readability} Others offered legibility-related suggestions and requests for customization about the visual attribute being modified. For instance, P6 suggested using \emph{``entirely different colors like RGB''} instead of shades of gray. P16 complained that \emph{``the lightest gray text was a bit difficult to read; I wish it could be a little darker.''} There are also font-size-related legibility concerns. For example, P12 thought \emph{``the text was too small and close together,''} and P14 wanted to see \emph{``an A+/A- icon by the side,''} so they could enlarge the font. 

\paragraph{Interactivity} P2 and P15 suggested additional features to interact with the grayed-out words. P15 proposed a slider to allow users to \emph{``customize the degrees of graying''} and P2 wished they could \emph{``put away the grayed out words''} entirely so they could focus on reading the words in black (a feature previously supported by \interactive{}). 
 
\subsubsection{\isac{} vs. \ngp{}: the Comparison Interview}

As described in Sec.~\ref{sec:mainstudy}, after completing all the reading tasks and reflecting on the three conditions of \isac{}, \wf{}, and \control{}, all the participants looked at the same passage shown twice, side by side, rendered with \isac{} and \ngp{}---one enforcing grammaticality at every minimum level of opacity and the other not. Fifteen (15) out of all 18 participants perceived a difference between the grammar-preserving and non-grammar-preserving renderings, although some could not specify exactly what the difference was. For instance, P2 noticed, \emph{``[\isac{}] seems to gray out longer chunks of text, while [\ngp{}] grays out a lot of single words.''} P11 mentioned, \emph{``In [\isac{}], the transition is much more natural. In [\ngp{}], honestly, I don't understand why certain words are in gray.''} P17 reported, \emph{``I actually didn't feel that much a difference, but I seemed to have an easier time reading in [\isac{}].''}

Half (9) of the 18 participants successfully identified, to varying degrees, that the key difference was in grammaticality. While some only sensed the difference, others were able to articulate specifically the grammatical errors in the \ngp{} case. For example, P9 successfully observed, \emph{``[\ngp{}] grays out many articles, prepositions, and other determiners, while [\isac{}] doesn't.''} Similarly, P18 elaborated, \emph{``[\ngp{}] grays out a lot of `the', `a' and `to', which is a little bit annoying to me. Those words may not carry much meaning, but they are still important to the structure of sentences.''}

As for user preferences, all 15 who perceived the difference between the two interfaces preferred \isac{} to \ngp{} because they felt it enabled them to achieve better comprehension and higher reading efficiency. For instance, P13 said, \emph{``I like [\isac{}] better. It just makes more sense to me. When I skipped the gray parts I still understood everything.''} P16 explained, \emph{``I prefer [\isac] because I can completely skip words in gray here but [in \ngp{}] I still have to read some of the gray text to understand what is going on.''} In summary, this part of the study provides evidence that \isac{}'s preservation of grammar at every level is key to the observed improvement in reading efficiency and user preference.

\section{Discussion}
These user studies demonstrate the benefits of Grammar-Preserving Text Saliency Modulation (\isac{}) for English reading comprehension. Participants responded positively to the chosen visual text attribute to be modified, i.e., text opacity, and especially strongly to the strategy by which text opacity was modulated, i.e., nested grammatical subsets of sentences that revealed layers of detail around the core of each sentence. One participant had reservations about the lack of predicted importance that the backend recursive extractive summarization process often assigned to transition words; this is evidence that the design goal concerning participants' ability to notice and recover in situations when they disagree with an automated judgement has been fulfilled, i.e., ``AI-resilience''. In spite of that keen observation, which was possible due to \isac{}'s design, there was a general consensus that \isac{}'s choices about which sections to gray out was superior to \wf{}, which is the nearest alternative method of text saliency modulation in the literature. Notably, the impact of the grammaticality enforced within \isac{}'s backend workflow was clearly perceived by most users; it garnered attention and praise in interviews when participants could see \isac{} side by side with its non-grammaticality-enforcing twin, \ngp{}. Multiple measures suggest that \isac{} enhances reading efficiency, overall user experience, and reduces the perceived difficulty of reading.

By far, our most compelling quantitative evidence are the gains in performance and decreases in task time when using \isac{}, compared to the controls. Participant performance on standardized test questions is less subjective than self-reported efficacy, which can be affected by social and cognitive biases, such as the lab setting, wanting to please the researchers or guessing the hypotheses. When the relative gains in efficiency are considered alongside findings from post-task surveys and qualitative feedback, there is strong evidence that \isac{}, as a visualization tool, supports faster and improved reading comprehension for English readers. 

The beauty of the \isac{} technique lies in its simplicity: at its core, all \isac{} does is change the visual saliency of words by adjusting their opacity. This preserves the integrity of the original text and minimizes ``ergonomic obtrusiveness''~\cite{yang2017hitext} while providing readers with a form of ``contextual cuing'' to arm them with ``incidental knowledge about global context'', which they can harness to better assign visual attention and memory when reading~\cite{healey2011attention}. By showing multiple levels of detail at once with successively less opacity, \isac{} empowers readers to freely choose their level of engagement with the material. By preserving the grammar at each level, \isac{} supports a coherent reading and skimming experience. 
The evidence from our user studies indicates that all our design goals were fulfilled.

Reflecting on the number of levels of opacity and their visual distinction, we encountered a tradeoff. We aimed for the least significant level to remain legible to ensure no loss of information, while also enabling clear differentiation among levels to allow readers to select and consistently engage with the level they consider the most suitable. However, the perceptibility of the differences among levels becomes challenging in complex sentences with many levels. Furthermore, according to Stevens's power law, people perceive changes in gray scale not linearly, but rather by a factor of approximately 0.5~\cite{munzner2014visualization}. For instance, a threefold increase in opacity might only be perceived as 1.5 times more significant, further complicating the differentiation of levels. This issue is reflected in feedback: some participants struggled to read the lightest gray text, while others had some difficulty discerning the various levels and understanding how they elucidate the relationships between different parts of sentences. 

Reflecting on the user experience, an intriguing transformation in reading patterns emerged from the feedback. Many participants pointed out, in one way or another, a two-step reading process that \isac{} interface seems to promote. Initially, readers focused on the darker, more salient text to grasp the primary narrative or theme of the passage. This 'overview' phrase of reading gave them a framework or scaffold of the content. Subsequently, they revisited the passage to delve into the grayed-out sections, filling in details where the questions were asked or their interest was piqued. This sequence resonates with efficient content absorption strategies highlighted in speed reading literature, where readers first capture the gist and then delve deeper~\cite{martiarini2013effects, abdelrahman2014effect}. The interface, therefore, may inadvertently facilitate this structured, layered reading approach, which might explain the improvement in reading efficiency and comprehension.

\section{Limitations and Future Work}
Reflecting on our technical approach, opting for an LLM-based backend enhanced the quality of the extractive summaries, but sacrificed speed and transparency. The black-box nature of LLMs reduces the transparency of the decisions they make, and their complexity slows down the system, potentially impacting future deployment of \isac{} to real-time reading scenarios. In particular, our choice of GPT-4 might limit the potential applications of our system due to data privacy concerns~\cite{khowaja2023chatgpt}; future work targeted at sensitive data should consider other open-source models that respect data privacy. In addition, the inherent non-determinism of LLMs can lead to variations in outputs for similar inputs, adding another layer of unpredictability in the LLM's responses. Although we partially mitigate this by requesting multiple responses and picking the best one algorithmically, our heuristic-based approach is not foolproof and may occasionally miss the most contextually relevant or coherent response. Despite these drawbacks, we still stick to an LLM-based approach because our primary focus at this stage remains optimizing the accuracy and relevance of text saliency modulation, which is currently best produced by LLM-based recursive extractive summarization at the paragraph level. As LLMs continue their current trend of advancement, we expect \isac{} to continue to improve in quality and speed, making it increasingly feasible to use---as one participant explicitly requested---as a Chrome extension.

Beyond the challenges posed by LLMs, our study also faces several other limitations. First, the limited sample size and sampling procedure could have skewed our conclusions due to a lack of diversity in participant background. Future evaluations of \isac{} should actively include a wider array of participants, such as younger or older age groups, users with varying educational backgrounds, and individuals from different cultural and linguistic contexts. These groups may encounter distinct challenges or exhibit different interaction patterns with \isac{}: age-related differences in technology adoption and comprehension skills, cultural variations in text interpretation, and educational disparities in reading abilities could all significantly impact the effectiveness of \isac{}. Expanding our understanding of these diverse user experiences is critical to a comprehensive understanding of the utility of \isac{} across a broader spectrum of users. Moreover, recent work~\cite{hendriksCHI23lbw} has identified needs of those with cognitive impairments, as well as possible directions for text tools to support them, such as helping readers \textit{prioritize} what to read. The evidence collected so far indicates that \isac{} may fulfill that need, but future evaluations of \isac{} should engage participants from that specific group to determine if \isac{} offers advantages for that community. 

Second, our measure of reading comprehension relied upon long passages from the GRE test, and how well \isac{} generalizes to other text styles and formats is yet unknown. This raises questions about the adaptability of \isac{} across various genres and complexities of text, such as technical manuals, legal documents, or everyday communication. Further, although our user study empirically evaluates the usability and usefulness of \isac{}, we rely solely on participants' accounts of their interactions to understand \emph{how} they used \isac{}, which could be subject to bias. Follow-up work could use eye-tracking studies to provide insights into how \isac{} shapes users' reading and skimming patterns. Finally, while we adhere to the guidelines provided by WCAG (Web Content Accessibility Guidelines) on contrast ratios of text, we acknowledge that modulating font opacity can make text less legible, and thus less accessible, especially to those with visual impairments.

Finally, we believe it is important to continue to explore the design space of \textit{AI-resilient} interfaces. Our understanding is that \isac{} is AI-resilient because, given that none of the original text is removed or rearranged, the errors of omission, hallucination, and misrepresentation instead show up as automated text attribute choices the reader disagrees with, and these automated choices are noticable, presented with all the necessary context for the reader to judge because: (1) Text attribute changes are always visible in the interface (i.e., no automated choice results in something hidden and therefore difficult to notice). (2) The reader is still looking at the original text so they have all the context they need to choose for themselves whether they agree with each automated choice or not (and what it implies about the text, e.g., whether that segment of text is particularly important or not). Generalizing this notion of AI-resiliency to additional tasks and domains is, we believe, important and exciting future work.

\bibliographystyle{ACM-Reference-Format}
\bibliography{sample-base}

\appendix
\section{Post-study Survey}
\label{appendix:survey}
\begin{table}[h!]
    \centering
    \begin{tabular}{l}
        \toprule
        \textbf{Question Statements} \\ \midrule
        Please rank the 3 interfaces from most to least helpful for answering the reading questions. \\ \midrule
        What did you like most about the interface you found the most helpful? [open-ended] \\ \midrule
        Are there any features missing that you'd like to see in the interface you found the most helpful? [open-ended] \\ \midrule
        I would like to use Reader-Blue to read online text of interest to me in the future. \\  \midrule
        I would like to use Reader-Green to read online text of interest to me in the future. \\ \midrule
        I would like to use Reader-Red to read online text of interest to me in the future. \\
        \bottomrule
    \end{tabular}
    \caption{Questions in the post-study survey. The last three ask participants to rate their agreement with them related to their reading experience on a 7-point Likert scale from ``Strongly Disagree'' (a score of 1) to ``Strongly Agree'' (a score of 7)}
    \label{tab:post_surey}
\end{table}

\section{\isac{} vs. \ngp{} in Comparison}
\label{appendix:comparison}
\begin{figure}[h!]
    \centering
    \includegraphics[width=0.9\textwidth]{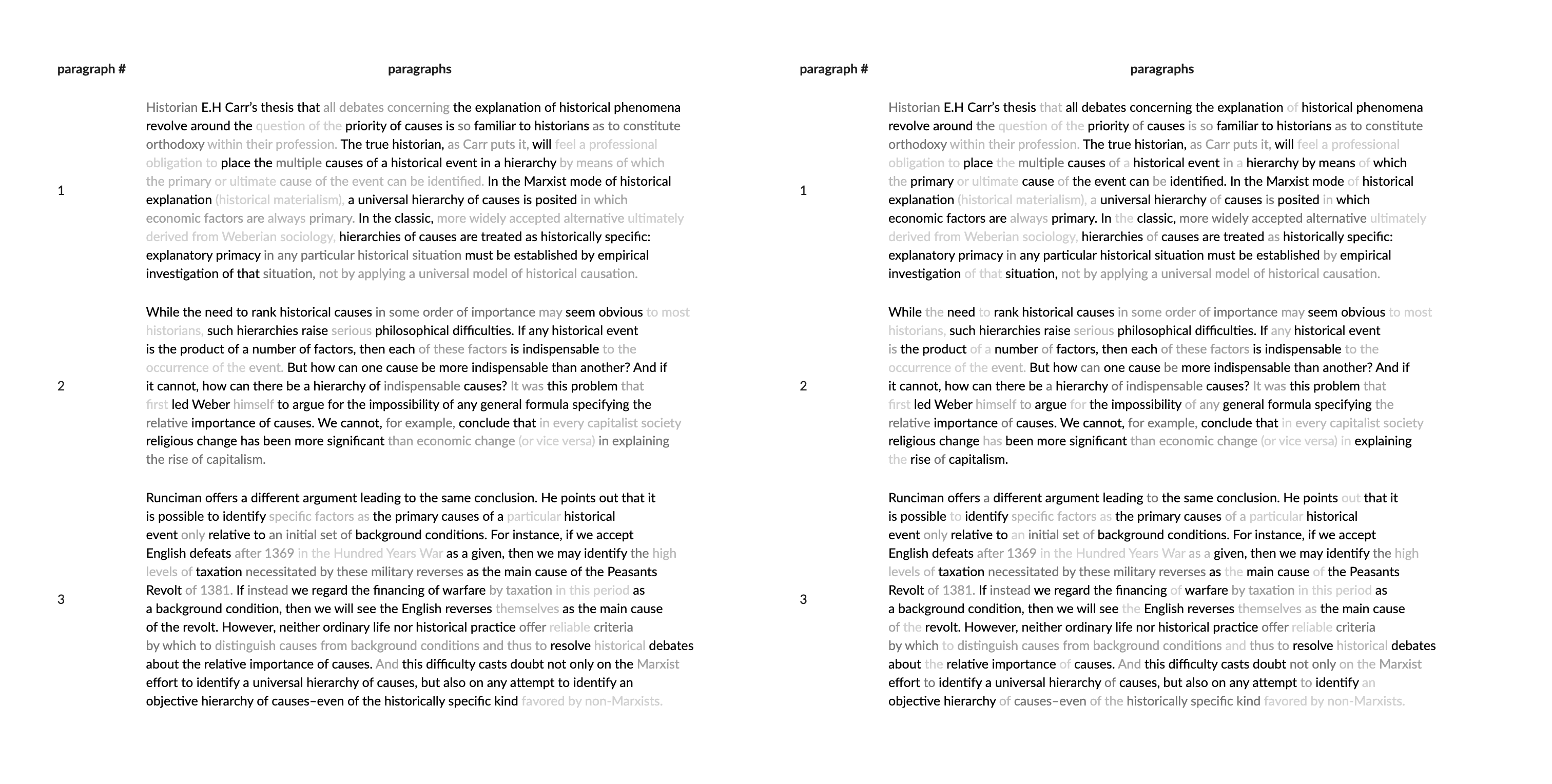}
    \caption{Side-by-side view of \isac{} (left) and \ngp{} (right) shown to a participant in one of the comparison interviews}
    \label{fig:comparison}
\end{figure}

\section{More Examples of \isac{}}
\label{appendix:examples}

\begin{figure}[h!]
    \centering
    \includegraphics[width=0.8\textwidth]{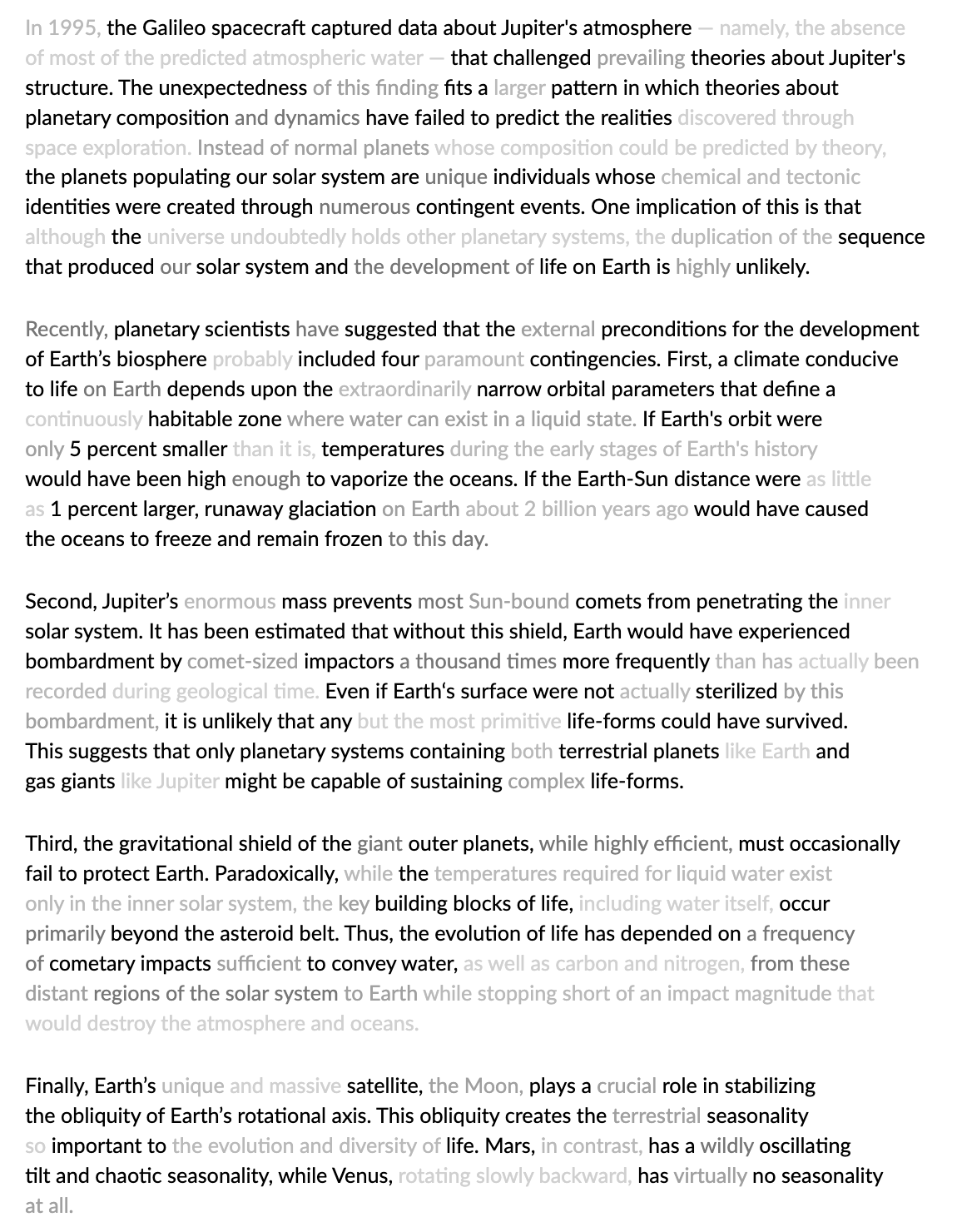}
    \caption{GRE Passage 1 rendered using \isac{}, as an additional example of how \isac{} works.}
\end{figure}

\begin{figure}[h!]
    \centering
    \includegraphics[width=0.8\textwidth]{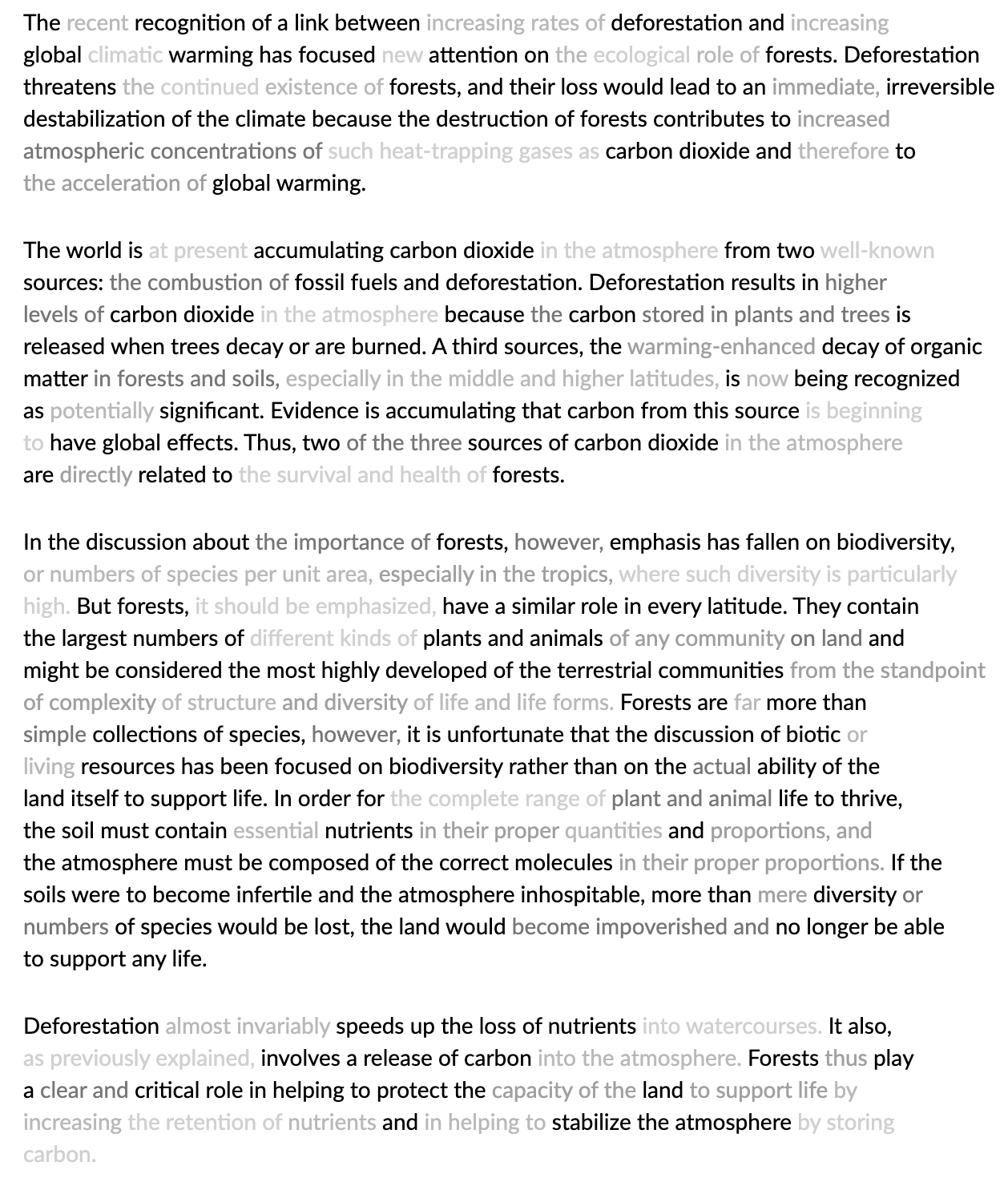}
    \caption{GRE Passage 2 rendered using \isac{}, as an additional example of how \isac{} works.}
\end{figure}

\begin{figure}[h!]
    \centering
    \includegraphics[width=0.8\textwidth]{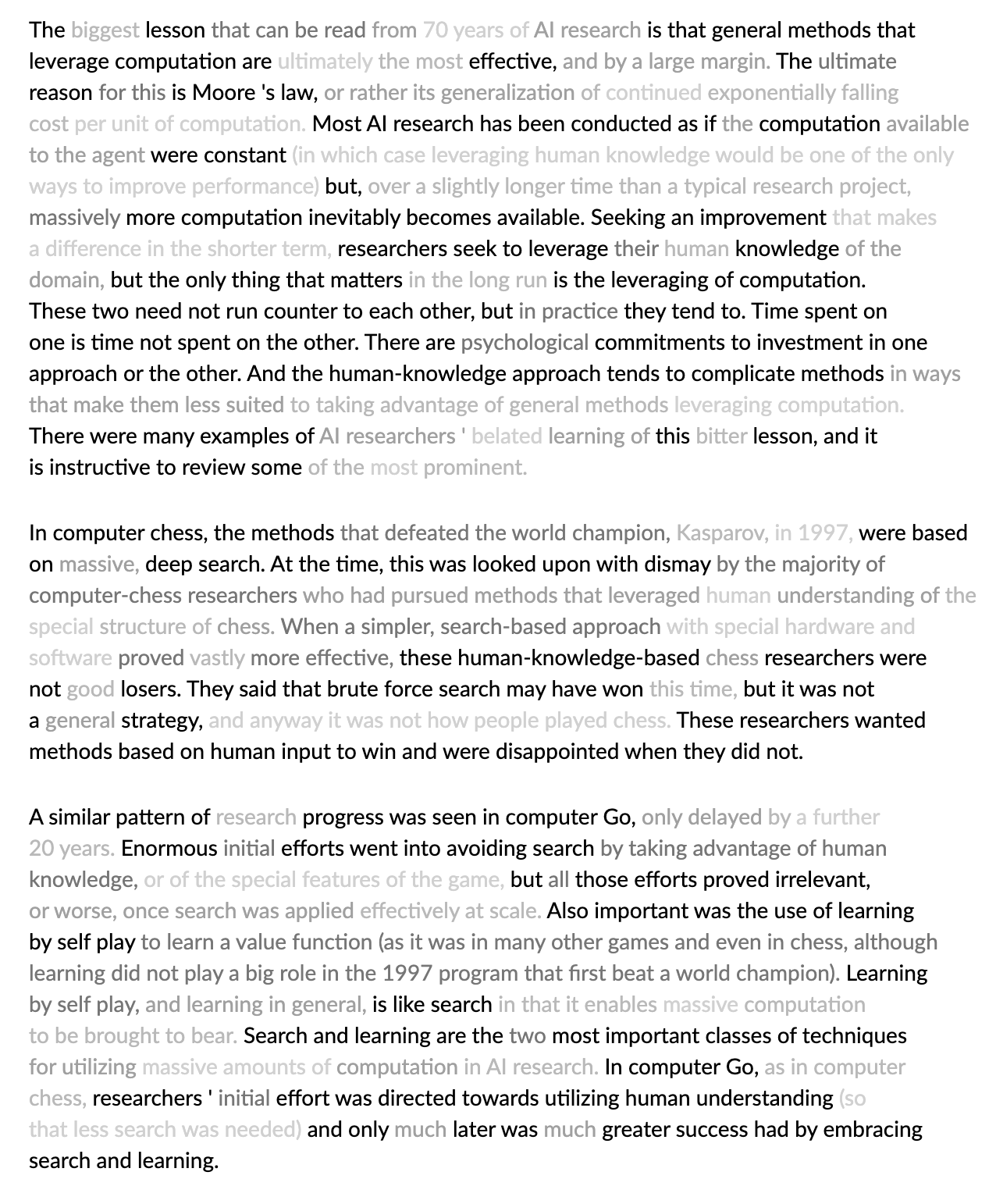}
    \caption{GRE Passage 3 rendered using \isac{}, as an additional example of how \isac{} works.}
\end{figure}

\end{document}